 \documentclass[preprint]{aastex}

\newcommand{\rs}{r_{_{\rm S}}}
\newcommand{\enrate}{\dot E}
\newcommand{\momrate}{\dot J}
\newcommand{\rc}{r_c}
\newcommand{\vff}{v_{\rm ff}}
\newcommand{\asym}{\doteq}
\newcommand{\ellK}{\ell_{\rm K}}
\newcommand{\OmegaK}{\Omega_{\rm K}}
\newcommand{\vc}{v_c}
\newcommand{\gapprox}{\lower.4ex\hbox{$\;\buildrel >\over{\scriptstyle\sim}\;$}}
\newcommand{\lapprox}{\lower.4ex\hbox{$\;\buildrel <\over{\scriptstyle\sim}\;$}}
\newcommand{\begeq}{\begin{equation}}
\newcommand{\fineq}{\end{equation}}
\def\vr{v}
\def\ellprime0{\ell'_0}

\slugcomment{accepted by \apj}

\shorttitle{Black Hole Inner Boundary Conditions}
\shortauthors{Becker \& Le}

\begin{document}

\title{INNER BOUNDARY CONDITIONS FOR ADVECTION-DOMINATED
ACCRETION ONTO BLACK HOLES \break}

\author{Peter A. Becker\altaffilmark{1}$^,$\altaffilmark{2}}

\and

\author{Truong Le\altaffilmark{3}}
\affil{Center for Earth Observing and Space Research, George Mason
University, Fairfax, VA 22030-4444, USA}

\vfil

\altaffiltext{1}{pbecker@gmu.edu}
\altaffiltext{2}{also Department of Physics and Astronomy,
George Mason University, Fairfax, VA 22030-4444, USA}
\altaffiltext{3}{tler@gmu.edu}

\begin{abstract}
The structure of the inner region of an advection-dominated accretion
disk around a nonrotating black hole is explored by applying asymptotic
analysis in the region just outside the event horizon. We assume that
the viscous transport is described by the standard Shakura-Sunyaev
prescription throughout the disk, including the inner region close to
the horizon. One of our goals is to explore the self-consistency of this
assumption by analyzing the causality of the viscous transport near the
black hole. The effects of general relativity are incorporated in an
approximate manner by utilizing a pseudo-Newtonian gravitational
potential. Analysis of the conservation equations yields unique
asymptotic forms for the behaviors of the radial inflow velocity, the
density, the sound speed, and the angular velocity. The specific
behaviors are determined by three quantities; namely, the accreted
specific energy, the accreted specific angular momentum, and the
accreted specific entropy. The additional requirement of passage through
a sonic point further constrains the problem, leaving only two free
parameters. Our detailed results confirm that the Shakura-Sunyaev
viscosity yields a well-behaved flow structure in the inner region that
satisfies the causality constraint. We also show that the velocity
distribution predicted by our pseudo-Newtonian model agrees with general
relativity in the vicinity of the horizon. The asymptotic expressions we
derive therefore yield useful physical insight into the structure of
advection-dominated disks, and they also provide convenient boundary
conditions for the development of global models via numerical
integration of the conservation equations. Although we focus here on
advection-dominated flows, the results we obtain are also applicable to
disks that lose matter and energy, provided the loss rates become
negligible close to the event horizon.

\end{abstract}



\section{INTRODUCTION}

The advection-dominated accretion flow (ADAF) model has received a great
deal of attention as a possible explanation for the dynamics occurring
in X-ray underluminous, radio-loud Active Galactic Nuclei (AGNs), which
are thought to contain hot accretion disks (e.g., Narayan \& Yi 1994,
1995). Briefly, these models describe the dynamics of gas fed onto a
black hole at very low (significantly sub-Eddington) accretion rates.
Gas accreting at such a low rate is quite tenuous, and therefore the
ion-electron Coulomb coupling timescale can exceed the timescale for
accretion. Since the ions absorb most of the energy dissipated via
viscosity, and the Coulomb coupling with the electrons is weak, the ions
achieve a nearly virial temperature ($T_i \sim 10^{12}\,$K) that greatly
exceeds the electron temperature ($T_e \sim 10^9\,$K), unless plasma
instabilities directly heat the electrons (Bisnovatyi-Kogan \& Lovelace
1997). The tenuous nature of the gas therefore severely limits the
radiative efficiency of the plasma, and consequently most of the thermal
energy dissipated by viscosity is advected into the black hole in the
form of hot protons, although outflows may carry some of this energy
away. The resulting X-ray luminosity is far below the Eddington limit.

In X-ray bright AGNs, the disk is thin and cool, and the accretion
proceeds in a radiatively efficient manner (Narayan 2002). This explains
the origin of the ``big blue bump'' in the spectra of typical Seyfert
galaxies, along with the common occurrence of broad emission lines,
apparently formed in the inner region between the last stable orbit and
the event horizon. The high temperatures in ADAF disks preclude the
formation of either the blue bump or the broad lines. From a theoretical
point of view, the reason a given object chooses one mode of accretion
over the other is not entirely clear. As Narayan (2002) points out, it
may be possible to learn a great deal about how the flow changes
character as a function of luminosity by studying transition objects
such as low-ionization nuclear emission regions (LINERs) and
low-luminosity active galactic nuclei (LLAGNs). Ptak et al. (1998)
suggest that the larger characteristic variability timescales observed
in these sources compared with brighter Seyfert galaxies may indicate
the presence of central ADAFs, which increase the size of the emission
region relative to more efficient, thinner disks. This idea is supported
by observations of the bright Seyfert 1 galaxy IC 4329A performed by
Done, Madejski, \& Zycki (2000) using {\it ASCA} and {\it RXTE}. They
note that the iron line in this source is not as broad as that detected
in more extreme cases such as MCG-6-30-15, where the cool disk
apparently extends down to the last stable orbit. Based upon this
observation, they conjecture that in IC 4329A the cool disk transitions
into a central ADAF outside the last stable orbit, and consequently the
inner region is too hot to produce the line emission. The physical
nature of the accretion disk may also vary as a function of time in
individual sources. For example, based on intensive {\it RXTE}
observations of the galactic black hole candidate J1550-564, Wilson \&
Done (2001) propose that transitions between the low/hard and high/soft
spectral states reflect the appearance and disappearance of an ADAF in
the inner region, perhaps as a consequence of changes in the accretion
rate. There are indications of similar behavior in some AGNs (e.g., Lu
\& Yu 1999).

These observations make it clear that in order to unravel the complex
global structure of the accretion disk, a complete understanding of the
physical properties of ADAFs close to the event horizon is essential.
One of the most important unresolved questions concerns the behavior of
the torque in the inner region, where the material begins to plunge into
the black hole. In the standard thin-disk model, it is usually assumed
that the stress vanishes at the marginally stable orbit (e.g., Frank,
King, \& Raine 1985). Others have used causality arguments to suggest
that the stress vanishes at the transonic point (e.g., Popham \& Narayan
1992). However, several authors have argued based on the results of
magnetohydrodynamical simulations that magnetic stresses are able to
remove angular momentum from the material in the plunging region (e.g.,
Reynolds \& Armitage 2001; Hawley \& Krolik 2001; Agol \& Krolik 2000;
Gammie 1999). The only model-independent statement that can be made with
absolute certainty is that the horizon itself cannot support a shear
stress. Hence we view this as the most conservative possible hypothesis.
The proposition that the viscous torque actually vanishes at some radius
outside the horizon is inevitably model-dependent. Following Narayan,
Kato, \& Honma (1997), Yuan (2001), and Yuan et al. (2000), we shall
therefore assume here that the torque vanishes at the horizon, and ask
whether self-consistent ADAF models with this property can be
constructed based upon the standard Shakura-Sunyaev viscosity
prescription. Our basic approach is to explore the associated disk
structure using rigorous asymptotic analysis. The validity of the
dynamical results is evaluated by examining the causality of the viscous
transport near the horizon, where all signals must propagate into the
black hole. We also compare the velocity distribution in the vicinity of
the horizon with the predictions of general relativity. Based upon these
considerations, we argue that a self-consistent Shakura-Sunyaev flow can
exist in the inner region, and we present detailed asymptotic solutions
for the disk structure.

The remainder of the paper is organized as follows. In \S~2 we briefly
review the sequence of models developed to describe ADAF disks. The
conservation equations for one-dimensional ADAF disks are discussed in
\S~3. The appropriate inner boundary conditions for ADAF disks are
derived in \S~4 by employing asymptotic analysis of the conservation
equations. In \S~5 the boundary conditions are used to obtain global
solutions for the physical quantities by numerically integrating the
conservation equations. The exact numerical solutions obtained are
compared with the asymptotic expressions and also with solutions
previously presented in the literature. The implications of our results
for the structure of advection-dominated accretion disks around black
holes are discussed in \S~6.

\section{DISK MODELS}

The structure of advection-dominated disks has been explored using a
variety of theoretical and computational approaches. The initial
one-dimensional, self-similar models developed by Narayan \& Yi (1994,
1995) incorporated Newtonian gravity. In later works, the assumption of
self-similarity was relaxed and new one-dimensional models were
developed based on a complete set of conservation equations (Narayan,
Kato, \& Honma 1997; Chen, Abramowicz, \& Lasota 1997). These models
utilized a pseudo-Newtonian gravitational potential in order to
approximate the effects of general relativity (Paczy\'nski \& Wiita
1980; Abramowicz, Calvani, \& Nobili 1980). Conservation of mass,
energy, and angular momentum is expressed using a set of coupled
differential equations that are integrated to obtain the inflow velocity
$\vr$, the sound speed $a$, and the angular velocity $\Omega$, as
functions of radius. The solution of these equations requires a
sufficient number of boundary conditions, imposed either at the sonic
point, the black hole horizon, or the outer radius of the computational
domain. The sonic point is a critical point for the flow, and therefore
the integration must be divided into two regions. One typically solves
the equations by starting at the sonic point with values that satisfy
the critical conditions, and then integrating the differential equations
inward towards the horizon, and outward toward larger radii, where the
disk is eventually expected to become thin and cool.

Because of the singularity at the event horizon introduced by the
pseudo-Newtonian potential, direct integration of the conservations
equations towards the horizon using an explicit method such as a
Runga-Kutta algorithm cannot be used to obtain smooth global solutions.
On the other hand, explicit integration {\it away} from the horizon
towards the critical point is stable, and therefore it represents a more
convenient approach to the problem, if the boundary conditions close to
the black hole event horizon can be specified. Another alternative is to
employ a relaxation algorithm that is based on iteration of the
numerical solution, with the goal of minimizing a global error parameter
(Press et al. 1986). Each of these methods requires the availability of
a suitable set of inner boundary conditions describing the physics of
the gas close to the event horizon. However, the boundary conditions
appropriate for this problem have not been presented previously in the
literature. Motivated by the lack of this crucial information, in this
paper we derive the exact asymptotic forms for the variation of the
physical quantities close to the event horizon by employing asymptotic
analysis based on the differential conservation equations. We shall
specialize to the case of one-dimensional, advection-dominated flow in
the pseudo-Newtonian potential. The asymptotic results will be used to
develop inner boundary conditions that facilitate the integration of the
conservation equations.

\section{CONSERVATION EQUATIONS}

We shall focus on the structure of steady, advection-dominated accretion
disks. These disks accrete at well below the Eddington rate, and are
therefore so tenuous that radiative cooling is inefficient. Hence
advection-dominated disks are nearly virially hot, and are essentially
collisionless. Moreover, the electrons and the protons are likely to
possess distinct temperatures due to the low rate of Coulomb
interactions between these species. In this situation, the protons
absorb most of the energy dissipated via viscosity, and consequently they
possess a much higher energy density than the electrons. Height-integrated
structure equations for such disks have been derived by Abramowicz
et al. (1988).

In our approach to modeling the disk structure, we will incorporate
the effects of general relativity in an approximate manner by expressing
the gravitational potential per unit mass using the pseudo-Newtonian
form (Paczy\'nski \& Wiita 1980)
\begeq
\Phi(r) = {- G \, M \over r - \rs} \ ,
\label{eq1}
\fineq
where $\rs = 2GM/c^2$ is the Schwarzschild radius for a black hole of
mass $M$. This potential gives correct results for the location of the
horizon, the radius of marginal stability, and the radius of the
marginally bound orbit around a nonrotating black hole (Paczy\'nski \&
Wiita 1980; Abramowicz, Calvani, \& Nobili 1980). Utilization of
equation~(\ref{eq1}) is convenient because it facilitates a
semi-classical approach to the problem that simplifies the analysis
considerably, while maintaining good agreement with fully relativistic
calculations. For this reason, the pseudo-Newtonian potential has been
adopted by a number of authors in their investigations of accretion onto
Schwarzschild black holes (e.g., Matsumoto et al. 1984; Abramowicz et
al. 1988; Chen et al. 1997; Narayan et al. 1997; Hawley \& Krolik 2001,
2002; Yuan 1999; Yuan et al. 2000; Reynolds \& Armitage 2001). We show
in Appendix~A that the relativistically correct energy equation for a
particle freely falling from rest at infinity in the Schwarzschild metric
can be written as
\begeq
{1 \over 2} \, v_r^2 + {1 \over 2} \, v_\varphi^2 + \Phi(r) = 0 \ ,
\label{eq1a}
\fineq
where $v_r$ and $v_\varphi$ denote the radial and azimuthal components of
the four-velocity, respectively, and $\Phi(r)$ is given by
equation~(\ref{eq1}). The Newtonian appearance of equation~(\ref{eq1a})
can be viewed as one of the primary motivations for introducing the
pseudo-Newtonian potential $\Phi(r)$, although one needs to keep
in mind that the dynamical quantities $v_r$ and $v_\varphi$
introduced in equation~(\ref{eq1a}) are four-velocities rather than
conventional velocities. We will return to this point later in our
discussion of the dynamical results.

\subsection{Transport Rates}

In our subsequent analysis, we will adopt the ``perfect ADAF'' approximation,
meaning that (for now) we shall completely neglect the escape of energy
and matter from the disk. Modifications associated with the relaxation of
this assumption will be discussed in \S~6.4. In the one-dimensional, steady
state ADAF scenario, three quantities are conserved in the flow, namely,
the accretion rate
\begeq
\dot M = 2 \pi r \cdot 2 H \cdot \rho \, \vr \ ,
\label{eq2}
\fineq
the angular momentum transport rate
\begeq
\momrate = \dot M \, r^2 \, \Omega - {\cal G} \ ,
\label{eq3}
\fineq
and the energy transport rate
\begeq
\enrate = - {\cal G}\, \Omega + \dot M \left( {1 \over 2}\,w^2
+ {1\over 2} \, \vr^2 + {P + U \over \rho} + \Phi\right)
\ ,
\label{eq4}
\fineq
where $\rho$ is the mass density, $\vr$ is the radial velocity (defined
to be positive for inflow), $\Omega$ is the angular velocity, $w = r \,
\Omega$ is the azimuthal velocity, $U$ is the internal energy density,
$P$ is the gas pressure, and ${\cal G}$ is the torque. All quantities
represent vertical averages over the disk half-thickness $H$. We shall
assume that the ratio of specifics heats, $\gamma \equiv (U + P) / U$,
maintains a constant value throughout the flow. Note that the transport
rates $\dot M$, $\momrate$, and $\enrate$ are all defined to be positive
for inflow. Since these quantities are conserved, they represent the
rates at which mass, angular momentum, and energy are swallowed by the
black hole. Athough we refer to $v$ and $w$ as ``velocities,'' we shall
see later that based on their asymptotic behavior close to the event
horizon, these quantities are actually more correctly interpreted as
four-velocities.

\subsection{Momentum Equations}

The expressions for the transport rates $\dot M$, $\momrate$, and
$\enrate$ are supplemented by equations describing the conservation
of the three components of momentum. In most of the one-dimensional
disk models, the thickness of the disk is computed using the assumption
of vertical hydrostatic equilibrium. While this assumption is probably
well-satisfied in the regions of the disk that have subsonic radial
velocities, it may not be very accurate near the horizon because the
flow is supersonic and practically in free-fall there. Despite this,
the hydrostatic assumption is routinely used to describe the entire
disk, including the supersonic region (e.g., Chen et al. 1997; Narayan
et al. 1997; Chen \& Taam 1993). Since one of our motivations is
to develop a consistent set of inner boundary conditions applicable
to ``standard'' ADAF models, we shall assume for now that the disk
half-thickness is given by the usual hydrostatic prescription,
\begeq
H(r) = {b_0 \, a \over \OmegaK} \ ,
\label{eq5}
\fineq
where $b_0$ is a dimensionless constant of order unity that depends
on the details of the vertical averaging (Abramowicz et al. 1988),
\begeq
a(r) \equiv \left({P \over \rho}\right)^{1/2}
\label{eq9}
\fineq
represents the isothermal sound speed, and $\Omega_{\rm K}(r)$ denotes
the Keplerian angular velocity of matter in a circular orbit at radius
$r$ in the pseudo-Newtonian potential (eq.~[\ref{eq1}]), defined by
\begeq
\Omega^2_{\rm K}(r) \equiv {GM \over r \, (r-\rs)^2}
= {1 \over r} \, {d\Phi \over dr} \ .
\label{eq10}
\fineq
In \S~6.2, we will discuss how our results would be modified if
the assumption of vertical hydrostatic equilibrium in the
supersonic region were replaced with radial free-fall.

In a steady state, the comoving radial acceleration rate in the frame of
the accreting gas is expressed by
\begeq
{Dv \over Dt} \equiv
\vr \, {d\vr \over dr} = - {1\over \rho} {dP \over dr}
- {d\Phi \over dr} + r \, \Omega^2 \ ,
\label{eq6}
\fineq
and angular momentum transport is treated by relating the torque
${\cal G}$ to the gradient of the angular velocity $\Omega$ using
the fundamental formula (e.g., Frank, King, \& Raine 1985)
\begeq
{\cal G} = - 4 \pi \, r \, H \, \rho \, \nu \, r^2 \,
{d\Omega \over dr} \ ,
\label{eq7}
\fineq
where $\nu$ is the kinematic viscosity.

\subsection{Viscosity and Torque}

We shall adopt the Shakura-Sunyaev (1973) prescription for the kinematic
viscosity,
\begeq
\nu = {\alpha \ a^2 \over \Omega_{\rm K}} \ ,
\label{eq8}
\fineq
where $\alpha$ is a constant. While this is certainly a reasonable
approach in the subsonic region of the flow, the validity of this
prescription in the supersonic region depends on the microphysical
mechanism responsible for generating the torque in a given situation. In
particular, if the angular momentum is transferred via turbulent blobs
of fluid, and the turbulence is subsonic, then the associated torque
should vanish in the supersonic region as a consequence of causality
considerations (e.g., Kato 1994; Narayan 1992; Popham \& Narayan 1992).
However, it is not clear a priori whether fluid turbulence, particles,
or magnetic fields (or some combination of these effects) transports the
angular momentum in ADAF disks. If the transport occurs via particles or
fields, then the causality argument mentioned above does not apply, and
torques can be generated even in the supersonic region of the flow
between the sonic point and the horizon, although the torque must
certainly vanish at the horizon itself in keeping with general
relativistic considerations. In fact, simulations performed by Reynolds
\& Armitage (2001) suggest that magnetic fields are able to transfer
angular momentum from material in the plunging region to material in the
outer disk. Following Narayan et al. 1997, we will therefore assume here
that the angular momentum is transferred by some generic
particle/magnetohydrodynamical mechanism, characterized by an effective
value for the adiabatic index $\gamma$ in the range $4/3 < \gamma <
5/3$. In this case, we can safely adopt the standard
$\alpha$-prescription for the kinematic viscosity $\nu$ given by
equation~(\ref{eq8}). The self-consistency of this approach will
be evaluated in \S~6.3, where the causality of the flow near the
horizon is examined.

\subsection{Entropy and Internal Energy}

Since we are neglecting the escape of energy from the disk, the comoving
rate of change of the internal energy density $U$ can be written in the
frame of the gas as
\begeq
{DU \over Dt} \equiv
- \vr \, {dU \over dr} = - \gamma \, {U \over \rho} \, \vr \, {d\rho \over
dr} + \dot U_{\rm viscous} \ ,
\label{eq11}
\fineq
where
\begeq
\dot U_{\rm viscous} = - \, {{\cal G} \over 4 \pi r H} {d\Omega \over dr}
= \rho \, \nu \, r^2 \left(d\Omega \over dr\right)^2
\label{eq12}
\fineq
is the viscous energy dissipation rate per unit volume. The negative signs
appearing in equation~(\ref{eq11}) reflect the fact that $\vr$ has been
defined to be positive for inflow. Combining equations~(\ref{eq8}), (\ref{eq11}),
and (\ref{eq12}), we can rewrite the internal energy equation as the entropy
equation
\begeq
\vr \, {d\over dr} \ln \left(U \over \rho^\gamma\right)
= - \, {\dot U_{\rm viscous} \over U}
= - \, {\alpha \, (\gamma-1) \, r^2 \over \Omega_{\rm K}}
\left(d\Omega\over dr\right)^2 \ .
\label{eq13}
\fineq
This equation demonstrates that the flow approaches a purely adiabatic
behavior ($U \propto \rho^\gamma$) wherever the viscous dissipation rate
$\dot U_{\rm viscous} / U$ vanishes. If the gas is in local thermodynamic
equilibrium, then the viscous heating is a quasi-static process, and in
this case the flow is {\it isentropic} wherever the dissipation vanishes.

We shall find it convenient to express the variation of the isothermal
sound speed, $a$, using the ``entropy function,''
\begeq
K(r) \equiv {r \, \vr \, a^{\gamma+1 \over \gamma-1} \over \Omega_{\rm K}}
\ .
\label{eq14}
\fineq
To understand the physical significance of $K$, we can combine
equations~(\ref{eq2}), (\ref{eq5}), (\ref{eq9}), and (\ref{eq14})
to show that
\begeq
K^{\gamma-1} \propto {U \over \rho^\gamma} \ ,
\label{eq15}
\fineq
which establishes that $K$ remains constant in regions of the flow
unaffected by dissipation. In particular, if the gas is in local thermodynamic
equilibrium, then we can use equation~(\ref{eq15}) to demonstrate that the
value of $K$ is related to the entropy per particle $S$ by (Reif 1965)
\begeq
S = k \, \ln K + c_0 \ ,
\label{eq17}
\fineq
where $c_0$ is a constant that depends only the composition of the
gas, but is independent of its state. Note that the relation between
$K$ and $S$ in equation~(\ref{eq17}) may be violated in an ADAF
because the gas is collisionless, and it is uncertain whether
collective processes can establish a Maxwell-Boltzmann distribution.
However, in any case $K$ itself is unambiguously defined by
equation~(\ref{eq14}). By comparing equations~(\ref{eq13}) and
(\ref{eq15}), we can show that the radial derivative of $K$ is
given by
\begeq
\vr \, {d \ln K \over dr} = 
- \, {\dot U_{\rm viscous} \over (\gamma-1) U}
= - \, {\alpha \, r^2 \over \Omega_{\rm K}}
\left(d\Omega \over dr\right)^2 \ .
\label{eq16}
\fineq
This result confirms that $K$ remains constant in regions of the flow
that are not subject to dissipation. We will employ equation~(\ref{eq16})
in \S~4.4, where we derive the asymptotic solution for the variation of the
entropy function close to the horizon.

\section{ASYMPTOTIC ANALYSIS}

The conservation equations presented in \S~3 can be solved as a coupled
set to determine the radial profiles of the physical quantities $\vr$,
$\Omega$, $a$, $\rho$, $P$, and $H$. Due to the divergence of the
pseudo-Newtonian potential as $r \to \rs$, the event horizon is a
regular singular point of the conservation equations governing the disk
structure. It is therefore possible to develop Frobenius-style
expansions of the physical quantities around the point $r = \rs$. Rather
than developing complete series solutions for the variables, we will
employ asymptotic analysis to determine the {\it dominant} behaviors as
$r \to \rs$. As we demonstrate below, this information can be used to
derive explicit boundary conditions applicable very close to the horizon.

\subsection{Stress Boundary Condition}

One of the fundamental boundary conditions for black hole accretion is
that the viscous shear stress $\Sigma$ must vanish as $r \to \rs$, because
particles at different radii become causally disconnected from each other.
The shear stress (force per unit area) is related to the torque ${\cal G}$
by
\begeq
\Sigma = - \rho \, \nu \, r \, {d\Omega \over dr}
= {{\cal G} \over 4 \pi r^2 H} \ ,
\label{eq18}
\fineq
and consequently the vanishing of the stress on the horizon implies that
${\cal G}=0$ there as well. We therefore conclude based on equation~(\ref{eq3})
that
\begeq
\lim_{r \to \rs} \Omega(r) \equiv \Omega_0
= {\momrate \over \dot M \, \rs^2}
\ ,
\label{eq19}
\fineq
which we can rewrite in terms of the azimuthal velocity $w = r \, \Omega$
as
\begeq
\lim_{r \to \rs} w = {\ell_0 \over \rs} \ ,
\label{eq19.3}
\fineq
where
\begeq
\ell_0 \equiv {\momrate \over \dot M}
\label{eq19.5}
\fineq
is the specific angular momentum of the material entering the black hole.
Hence $w$ approaches a finite value at the event horizon.
Let us consider the physical implications of this result. Recall that the
azimuthal velocity $v^{\hat \varphi}$ of a freely-falling particle as
measured by a stationary observer outside a Schwarzschild black hole
{\it vanishes} at the horizon (Shapiro \& Teukolsky 1983). Hence,
$w$ cannot represent the true azimuthal velocity measured by a static
observer in the region close to the horizon. In Appendix~A, we demonstrate
that $w$ is actually equal to the azimuthal component of the particle's
{\it four-velocity}, $v_\varphi$, and consequently $w$ possesses a finite
value at the horizon.

\subsection{Radial Velocity}

Our result for the asymptotic variation of $\Omega$ close to the horizon
(eq.~[\ref{eq19}]) can be used to derive the corresponding behavior of
the radial inflow velocity $v$ as $r \to \rs$. By using
equation~(\ref{eq3}) to eliminate the torque in equation~(\ref{eq4}), we
can rewrite the energy transport equation as
\begeq
\enrate = \momrate \, \Omega  + \dot M \left({1 \over 2} \, \vr^2
- {1 \over 2} \, r^2 \Omega^2 + {\gamma \over \gamma-1} \, a^2
- {GM \over r - \rs} \right) \ ,
\label{eq20}
\fineq
where we have also substituted for the potential $\Phi$ using
equation~\ref{eq1}. The flow into the black hole must be supersonic
at the horizon since the radial velocity approaches the speed of
light there. Hence $\vr \gg a$ as $r \to \rs$. Since we have determined
that the angular velocity $\Omega$ approaches a finite value at the
horizon, we can conclude based on equation~(\ref{eq20}) that the radial
velocity approaches the free-fall velocity $\vff(r)$, i.e.,
\begeq
\lim_{r \to \rs} {\vr(r) \over \vff(r)} = 1 \ , \ \ \ \ \ \ 
\vff(r) \equiv \left(2 \, GM \over r - \rs\right)^{1/2} \ .
\label{eq21}
\fineq
This implies that close to the horizon, $\vr$ formally exceeds $c$, and
therefore it cannot represent the actual radial velocity $v^{\hat r}$
measured by a static local observer in the Schwarzschild metric. In
Appendix~A we demonstrate that close to the horizon, $v$ is actually
equal to the radial component of the four-velocity, $v_r$, for a freely
falling particle.

\subsection{Angular Momentum and Torque}

The vanishing of the stress at the horizon ensures that the disk
experiences differential rotation with $d\Omega / dr \le 0$ at all
radii. We can use this observation along with the requirement that
$\Omega \to \Omega_0$ as $r \to \rs$ to develop the leading-order
behavior of the Frobenius expansion for $\Omega(r)$ about $r = \rs$.
Since $r = \rs$ is a regular singular point of the conservation
equations, we can in general write the asymptotic behavior of
$\Omega(r)$ close to the horizon as
\begeq
\Omega(r) \asym \Omega_0 - A \, (r - \rs)^q \ ,
\label{eq22}
\fineq
where $A$ and $q$ are positive constants and we will use the symbol
``$\asym$'' to denote asymptotic equality at the horizon. Equation~(\ref{eq22})
is the simplest form that satisfies the requirements that $\Omega \to
\Omega_0$ and $d\Omega / dr \le 0$ as $r \to \rs$. The right-hand side
of equation~(\ref{eq22}) represents the first two terms of the Frobenius
expansion for $\Omega(r)$, and $q$ is the exponent of the solution about
$r = \rs$ (Boyce \& DiPrima 1977). We can constrain the value of $q$ by
examining the associated variation of the specific angular momentum,
$\ell \equiv r^2 \, \Omega$. Differentiation of $\ell$ with respect to
radius yields
\begeq
{d \ell \over dr} = 2 \, r \, \Omega + r^2 \, {d\Omega \over dr}
\ .
\label{eq23}
\fineq
Substituting for $\Omega$ using equation~(\ref{eq22}), we obtain the
asymptotic relation
\begeq
{d\ell \over dr} \asym 2 \, r \, \Omega_0 - 2 \, A \, r (r - \rs)^q
- A \, q \, r^2  (r - \rs)^{q-1} \ .
\label{eq24}
\fineq
Now, according to equation~(\ref{eq3}), $\ell = (\momrate + {\cal G})/\dot M$,
and therefore $d\ell/dr \ge 0$ at the horizon since the torque ${\cal G}$
vanishes there. This in turn implies that $d\ell/dr$ has a finite, positive
value at the horizon. Based on this constraint, we conclude that $q \ge 1$,
since otherwise $d\ell /dr$ would diverge to negative infinity at the
horizon.

The conclusion that $q \ge 1$ implies that $d\Omega/dr$ approaches
a {\it finite value} as $r \to \rs$. Furthermore, since $\vr$ and
$\Omega_{\rm K}$ each diverge as $r \to \rs$, we can demonstrate based
on equation~(\ref{eq13}) that the flow displays a purely adiabatic behavior
close to the horizon (i.e., $U \propto \rho^\gamma$). This explicitly
confirms that the dissipation vanishes at the horizon, which is of
course intuitively obvious since the stress vanishes there. According to
equation~(\ref{eq16}), the entropy function $K$ consequently approaches a
finite value at the horizon, i.e.,
\begeq
\lim_{r \to \rs} K(r) \equiv K_0 \ ,
\label{eq25}
\fineq
where $k \ln K_0 + c_0$ represents the specific entropy of the particles
entering the black hole (see eq~[\ref{eq17}]).

We can build on our previous conclusions to further explore the asymptotic
behavior of the specific angular momentum close to the horizon. Combining
equations~(\ref{eq2}), (\ref{eq3}), (\ref{eq7}), (\ref{eq8}), and
(\ref{eq19.5}), we can express the radial derivative of $\Omega$ as
\begeq
{d\Omega \over dr} = - \, {\vr \, \Omega_{\rm K} \, (\ell - \ell_0)
\over \alpha \, r^2 \, a^2} \ .
\label{eq26}
\fineq
Using this result to substitute for $d\Omega/dr$ in equation~(\ref{eq23})
yields a differential equation for $\ell$,
\begeq
{d \ell \over dr} = {2 \, \ell \over r} - {\vr \, \OmegaK \,
(\ell - \ell_0) \over \alpha \, a^2} \ .
\label{eq27}
\fineq
As was pointed out earlier, $\ell = (\momrate + {\cal G})/\dot M$,
and therefore $d\ell/dr \ge 0$ at the horizon since ${\cal G} \to 0$
as $r \to \rs$. It follows that the local behavior of $\ell$ close to
the horizon must be of the general form
\begeq
\ell(r) \asym \ell_0 + B (r-\rs)^\beta \ ,
\label{eq28}
\fineq
where $B$ and $\beta$ are positive constants. This represents the
leading behavior of the Frobenius expansion for $\ell(r)$ about $r =
\rs$, and $\beta$ is the exponent of the solution. Paczy\'nski \& Wiita
(1980) imposed equation~(\ref{eq28}) as an ad hoc expression for the
{\it global} variation of the specific angular momentum $\ell(r)$, whereas
we employ it only in the asymptotic limit, where its validity is fully
supported by the conservation equations. In fact, we shall see that this
is not a very accurate expression for $\ell$ far from the event horizon
in our application. Using equation~(\ref{eq28}) to substitute for $\ell$
on the right-hand side of equation~(\ref{eq27}) yields in the limit $r
\to \rs$
\begeq
\lim_{r \to \rs} {\vr \, \OmegaK \over \alpha \, a^2} \, B \, (r-\rs)^\beta
= {2 \, \ell_0 \over \rs} - \ellprime0 \ ,
\label{eq29}
\fineq
where
\begeq
\ellprime0 \equiv \lim_{r \to \rs} {d\ell \over dr} \ .
\label{eq30}
\fineq
The constants $B$, $\beta$, and $\ellprime0$ are determined as follows.
To evaluate the limit on the left-hand side of equation~(\ref{eq29}), we
substitute for $a$ using equation~(\ref{eq14}) and set $K = K_0$ at the
black hole horizon. We have already determined that the gas approaches
radial free-fall as $r \to \rs$, i.e., $\vr \to [2GM/(r-\rs)]^{1/2}$.
Using this information, we obtain
\begeq
\lim_{r \to \rs} {2^{1/2} \, GM \over \alpha \, r^{1/2} (r-\rs)^{3/2}}
\left[2 \, r^3 \, (r-\rs) \over K_0^2 \right]^{\gamma-1 \over \gamma+1}
B \, (r-\rs)^\beta
= {2 \, \ell_0 \over \rs} - \ellprime0 \ .
\label{eq31}
\fineq
In order to obtain a constant value on the left-hand side in the limit
$r \to\rs$, we must require that the exponents of $(r-\rs)$ add to zero.
This yields the result
\begeq
\beta = {\gamma+5 \over 2 \, (\gamma+1)} \ .
\label{eq32}
\fineq

In Table~1 we list the values of $\beta$ obtained for several values of
$\gamma$. Note that for $\gamma$ is the range $4/3 \le \gamma \le 5/3$,
we find that $1.36 \ge \beta \ge 1.25$. Hence $\beta$ exceeds unity for
any physically acceptable equation of state. Using equation~(\ref{eq28})
to evaluate $d\ell/dr$ in the limit $r \to \rs$ therefore yields
\begeq
\ellprime0 = \lim_{r \to \rs} B \, \beta (r-\rs)^{\beta-1} = 0 \ .
\label{eq33}
\fineq
Hence we have proven that $d \ell/dr = 0$ at the black hole event horizon.
By balancing the values on the two sides of equation~(\ref{eq31}) in the limit
$r \to \rs$, it is straightforward to show that the constant $B$ is given
by
\begeq
B = {\alpha \, \ell_0 \over GM} \left(2 \over \rs \right)^{1/2}
\left(K_0^2 \over 2 \, \rs^3 \right)^{(\gamma-1) / (\gamma+1)} \ .
\label{eq34}
\fineq
Combining equations~(\ref{eq28}), (\ref{eq32}), and (\ref{eq34}), we can
express the asymptotic solution for $\ell$ near the horizon as
\begeq
\ell(r) \asym \ell_0 +
{\alpha \, \ell_0 \over GM} \left(2 \over \rs \right)^{1/2}
\left(K_0^2 \over 2 \, \rs^3\right)^{\gamma-1 \over \gamma+1}
(r - \rs)^{\gamma + 5 \over 2 \gamma + 2} \ .
\label{eq35}
\fineq
This relation gives the dominant asymptotic behavior of the specific
angular momentum as a function of $r$ for arbitrary values of $\alpha$,
$\gamma$, $\ell_0$, and $K_0$. It is interesting to note that although
Paczy\'nski \& Wiita (1980) arbitrarily imposed equation~(\ref{eq28})
as a global expression for $\ell(r)$, our numerical results for $\beta$
are relatively close to the values they obtain.

Our asymptotic result for $\ell(r)$ has two important implications.
First, since we have found that $d \ell/dr = 0$ at the event horizon,
it follows from consideration of equation~(\ref{eq24}) that $q = 1$ and
$A = 2 \, \Omega_0 / \rs$. Referring to equation~(\ref{eq22}), we conclude
that the asymptotic solution for $\Omega(r)$ is therefore given by
\begeq
\Omega(r) \asym \Omega_0 - {2 \, \Omega_0 \over \rs} \, (r - \rs)
\label{eq36}
\fineq
in the vicinity of the horizon. Note in particular that $d\Omega/dr =
-2 \, \Omega_0 /\rs$ at the horizon, in contradiction to Narayan et al.
(1997), who erroneously stated that $d\Omega/dr = 0$ there. Second, since
the torque ${\cal G}$ is linearly related to $\ell$ via ${\cal G}
= \dot M \ell - \momrate = \dot M (\ell - \ell_0)$, we find that the
asymptotic variation of the torque is given by
\begeq
{\cal G}(r) \asym
{\alpha \, \ell_0 \dot M \over GM} \left(2 \over \rs \right)^{1/2}
\left(K_0^2 \over 2 \, \rs^3\right)^{\gamma-1 \over \gamma+1}
(r - \rs)^{\gamma + 5 \over 2 \gamma + 2} \ .
\label{eq37}
\fineq
Based on this expression, we conclude that {\it the radial derivative of
the torque vanishes at the horizon}, i.e.,
\begeq
\lim_{r \to \rs} {d {\cal G} \over dr} =0 \ .
\label{eq38}
\fineq
This completely new boundary condition is one of the main results
of the paper. The vanishing of the derivative $d{\cal G}/dr$
at the horizon supplements the well-known boundary condition
${\cal G} = 0$. The physical interpretation of this new boundary
condition and its effect on the structure of the global flow solutions
will be discussed in \S\S~5 and 6.

\subsection{Entropy}

Our insights regarding the asymptotic behaviors of $\ell$, $\Omega$,
$a$, and $\vr$ can be combined to ascertain the asymptotic variation
of the entropy function $K(r)$ close to the horizon. Substituting into
equation~(\ref{eq16}) for $d\ln K/dr$ using the asymptotic free-fall
velocity (eq.~[\ref{eq21}]) along with equation~(\ref{eq10}) for
$\Omega_{\rm K}$ gives the leading behavior
\begeq
{d\ln K \over dr} \asym - {4 \, \alpha \, \Omega_0^2 \over GM}
\left(\rs \over 2 \right)^{1/2} \left(r - \rs\right)^{3/2}
\label{eq39}
\fineq
near to the horizon, where we have also used the fact that $d\Omega/dr=
-2 \, \Omega_0/\rs$ at $r=\rs$. Integration of equation~(\ref{eq39}) with
respect to radius yields for the asymptotic behavior of $K(r)$ the solution
\begeq
K(r) \asym K_0 \left[1 - {8 \, \alpha \, \Omega_0^2 \over 5 \,  GM}
\left(\rs \over 2 \right)^{1/2} (r-\rs)^{5/2} \right]
\label{eq40}
\fineq
in the vicinity of the horizon. Note that $K \to K_0$ at the horizon as
required, although the radial dependence is very weak, reflecting the
gradual disappearance of viscous dissipation as $r \to \rs$. The flow
is therefore essentially isentropic close to the horizon.

\subsection{Sound Speed and Inflow Velocity}

We have shown that $\vr$ approaches the radial free-fall velocity $\vff
= [2GM/(r-\rs)]^{1/2}$ near the horizon. While this result is certainly
valid in the limit $r \to \rs$, we may strive to obtain a more precise
formula for $\vr$ by employing the insights we have obtained in our
study of the local variations of the specific angular momentum $\ell(r)
=r^2 \, \Omega(r)$ and the entropy function $K(r)$ close to the horizon.
Based on equation~(\ref{eq20}), we can express the energy per unit mass
transported through the disk as
\begeq
\epsilon_0 \equiv {\enrate \over \dot M} =
{\vr^2 \over 2} - {\ell^2 \over 2 \, r^2} + {\ell_0 \ell \over r^2}
+ {\gamma \over \gamma-1} \, a^2 - {GM \over r-\rs} \ ,
\label{eq41}
\fineq
where $\ell_0 \equiv \momrate / \dot M$ is the accreted specific angular
momentum. Since $\enrate$ and $\dot M$ are conserved, it follows that
$\epsilon_0$ represents the energy per unit mass swallowed by the black
hole.

Close to the horizon, the viscous dissipation vanishes, and $K \to K_0$.
We can therefore use equation~(\ref{eq14}) to express the variation of
the isothermal sound speed $a$ in the vicinity of the horizon as
\begeq
a(r) \asym \left(K_0 \, \Omega_{\rm K} \over r \, \vr \right)
^{\gamma-1 \over \gamma+1} \ .
\label{eq42}
\fineq
We also know that $\ell \to \ell_0$ as $r \to \rs$. Using this condition
along with equation~(\ref{eq42}), we can rewrite equation~(\ref{eq41})
as the asymptotic expression
\begeq
\epsilon_0 \asym
{\vr^2 \over 2} + {\ell_0^2 \over 2 \, r^2}
+ {\gamma \over \gamma-1} \, \left(r^2 \vr^2 \over K_0^2 \, \Omega_{\rm K}^2
\right)^{1 - \gamma \over 1 + \gamma} - {GM \over r-\rs} \ .
\label{eq43}
\fineq
This nonlinear relation governs the variation of the inflow velocity
$v(r)$ close to the horizon. In general,
it must be solved numerically to determine $\vr$ for given values of $\ell_0$,
$K_0$, $\epsilon_0$, and $r$. However, as an alternative to numerical
root-finding, we can seek an asymptotic, analytical solution for $\vr(r)$
by expanding equation~(\ref{eq43}) in the small parameter $g(r)$, defined by
\begeq
\vr^2(r) \equiv \vff^2(r) \left[1 + g(r) \right] \ .
\label{eq44}
\fineq
Very close to the horizon, the velocity approaches free-fall, and therefore
we must have $g(r) \to 0$ as $r \to \rs$. Using equation~(\ref{eq44}) to
substitute for $\vr^2$, we can linearize the factor in parentheses in
equation~(\ref{eq43}) to obtain
\begeq
\epsilon_0 \asym
{g \, \vff^2 \over 2} + {\ell_0^2 \over 2 \, r^2}
- {\gamma \over \gamma + 1} \, \left(r^2 \vff^2 \over K_0^2 \, \Omega_{\rm K}^2
\right)^{1 - \gamma \over 1 + \gamma} \left({1 + \gamma \over 1 - \gamma} +
g \right) \ .
\label{eq44.5}
\fineq
Solving this equation for $g(r)$ yields the asymptotic result
\begeq
g(r) \asym {2 \, \epsilon_0 \, r^2 - \ell_0^2 - (\gamma+1) \, f(r)
\over r^2 \, \vff^2(r) - (\gamma-1) \, f(r)} \ ,
\label{eq45}
\fineq
where
\begeq
f(r) \equiv {2 \, \gamma \, r^2 \over \gamma^2-1} \left[K_0^2 \over 2 \,
r^3 (r-\rs) \right]^{\gamma-1 \over \gamma+1} \ .
\label{eq46}
\fineq
Note that the function $f(r)$ diverges as $r \to \rs$, but it does so much
more slowly than $\vff^2(r)$, and consequently $g(r) \to 0$ at the horizon
as required. Equations~(\ref{eq44}), (\ref{eq45}), and (\ref{eq46}) provide
a useful asymptotic representation for the inflow velocity $\vr(r)$ that
describes the first-order correction to purely free-fall behavior close
to the horizon.

By combining equations~(\ref{eq10}), (\ref{eq42}), and (\ref{eq44}), we can
show that the asymptotic solution for $a(r)$ is given by
\begeq
a(r) \asym \left(K_0^2 \over 2 \, r^3 \right)^{-\sigma}
\left[1+g(r)\right]^\sigma \, (r - \rs)^\sigma \ ,
\label{eq47}
\fineq
where
\begeq
\sigma \equiv {1-\gamma \over 2(\gamma+1)} \ .
\label{eq47.5}
\fineq
The dominant behavior as $r \to \rs$ is $a \propto (r-\rs)^\sigma$.
The exponent $\sigma$ is negative, and therefore $a$ diverges
at the horizon, albeit much more slowly than $\vr$, which approaches
$\vff \propto (r-\rs)^{-1/2}$.

We can also easily determine the leading behavior of the disk
half-thickness $H(r)$ close to the horizon by combining the hydrostatic
relation $H = b_0 \, a / \OmegaK$ with equations~(\ref{eq10}) and
(\ref{eq47}), which yields
\begeq
H(r) \asym {b_0 \over c} \left(K_0^2 \over 2 \, r^3 \right)^{-\sigma}
\left[1+g(r)\right]^\sigma \left(2 \, r \over \rs \right)^{1/2}
(r - \rs)^\delta
\ , \label{eq47.7}
\fineq
where $c$ is the speed of light and
\begeq
\delta \equiv {\gamma + 3 \over 2(\gamma+1)} \ .
\label{eq47.9}
\fineq
As $r \to \rs$, the dominant behavior is $H(r) \propto (r-\rs)^\delta$.
Since the flow becomes adiabatic close to the horizon, we can use
equations~(\ref{eq9}) and (\ref{eq47}) to show that the dominant
asymptotic forms for the pressure $P$ and the mass density $\rho$ are
given by
\begeq
P(r) \propto (r - \rs)^\lambda \ , \ \ \ \ \ \ \ 
\rho(r) \propto (r - \rs)^\eta \ ,
\label{eq47.91}
\fineq
where
\begeq
\lambda \equiv - \, {\gamma \over \gamma+1} \ , \ \ \ \ \ \ \ 
\eta \equiv - \, {1 \over \gamma+1}
\ .
\label{eq47.92}
\fineq
Table~1 includes values for the exponents $\sigma$, $\delta$, $\lambda$,
and $\eta$ obtained for several different values of the adiabatic index
$\gamma$. Note that $\delta \sim 1$, indicating that $H$ is roughly
proportional to $r - \rs$. Hence the disk has zero thickness (i.e., a
cusp) at $r = \rs$, reflecting the fact that the gas pressure is unable
to support the disk against the strong gradient of the gravitational
potential as the matter approaches the horizon. This particular behavior
is a manifestation of the assumption of vertical hydrostatic
equilibrium. In \S~6.2 we explore the consequences of replacing this
assumption with radial free-fall close to the horizon.

Equations~(\ref{eq44}) and (\ref{eq47}) provide extremely accurate
asymptotic solutions for $\vr(r)$ and $a(r)$, respectively. The accuracy
of these solutions will be demonstrated in \S~5 by comparing them with
exact solutions obtained by integrating numerically the differential
conservation equations. Taken together, equations~(\ref{eq35}),
(\ref{eq40}), (\ref{eq44}), and (\ref{eq47}) completely determine the
nature of the flow close to the horizon in terms of the three free
parameters $\ell_0$, $K_0$, and $\epsilon_0$, which describe the
specific angular momentum, the specific entropy, and the specific energy
of the particles entering the black hole, respectively. As discussed in
\S~5, the requirement of smooth passage through a critical point imposes
an additional constraint that effectively reduces the number of free
parameters from three to two. The asymptotic results we have derived in
this section can be used to define boundary conditions applicable close
to the event horizon that serve as the basis for numerical simulations
of the global structure of advection-dominated disks. In \S~5 we perform
global disk structure calculations and compare the numerical solutions
obtained with our asymptotic expressions in the vicinity of the horizon.

\section{GLOBAL FLOW SOLUTIONS}

The various differential and algebraic conservation equations may be
solved numerically to determine the profiles of $\vr$, $a$, and
$\ell = r^2 \, \Omega$. Several methods are available to solve the coupled
system of equations, such as explicit integration using a Runga-Kutta
solver, or the utilization of a global, iterative relaxation method
(e.g., Press et al. 1986). Each of these techniques requires the
imposition of boundary conditions at the edges of the computational
domain. The new asymptotic relations derived in \S~4 can be applied
at a radius just outside the horizon to provide the inner boundary
conditions at the black hole event horizon needed to compute global
solutions for the disk structure. Before attempting to solve the
computational problem to determine the disk properties, it is worthwhile
to review the critical nature of the conservation equations.

\subsection{Dynamical Equation and Critical Conditions}

Black hole accretion flows are in general transonic, and therefore the
computational domain can be broken into two regions, one above the sonic
radius and one below it. Successful global solutions must pass smoothly
through the sonic point, which is a critical point for the flow. In
order to explore the critical nature of the flow, it is convenient to
derive a dynamical equation based on the mass, momentum, and energy
conservation equations. As a preliminary step, we can use
equations~(\ref{eq2}), (\ref{eq5}), and (\ref{eq9}) to express the
density $\rho$ in terms of $\vr$, $r$, and $P$ as
\begeq
\rho = \left( \dot M \, \OmegaK \over 4 \, \pi \, r \, \vr \,
b_0 \right)^2 {1 \over P} \ .
\label{eq48}
\fineq
Using this relation to eliminate $\rho$ in the entropy equation~(\ref{eq13}),
we can derive an equation for the pressure derivative $dP/dr$. The result
obtained is
\begeq
{\gamma+1 \over 2 \, \gamma} \, {d\ln P \over dr} + {1 \over r} + {d\ln v \over dr}
- {d\ln\OmegaK \over dr} = - {\alpha (\gamma-1) r^2 \over 2 \, \gamma v \OmegaK}
\left(d\Omega \over dr\right)^2 \ .
\label{eq48.5}
\fineq
This can be used to eliminate the pressure derivative in the radial
momentum equation~(\ref{eq6}) to yield the dynamical equation
\begeq
\left({\vr^2 \over a^2} - {2 \, \gamma \over \gamma+1}\right)
{d\ln \vr \over dr} = {\ell^2 - \ellK^2 \over a^2 \, r^3}
+ {2 \, \gamma \over \gamma+1} \left({3 \over r} - {d\ln\ellK
\over dr}\right) - \left(\gamma-1 \over \gamma+1\right) {\vr \,
\ellK \, (\ell - \ell_0)^2 \over \alpha \, a^4 \, r^4} \ ,
\label{eq49}
\fineq
where
\begeq
\ellK(r) \equiv r^2 \, \OmegaK(r) = {(G M)^{1/2} \, r^{3/2} \over
r - \rs}
\label{eq50}
\fineq
denotes the specific angular momentum of particles in circular,
Keplerian orbits at radius $r$, and we have also used
equations~(\ref{eq10}) and (\ref{eq26}). Equation~(\ref{eq49}) agrees
with equation~(2.16) of Narayan et al. (1997). Global flow solutions can
be obtained by integrating simultaneously the two coupled differential
equations~(\ref{eq27}) and (\ref{eq49}), which govern the functions
$\ell(r)$ and $\vr(r)$, respectively. This is similar to the procedure
followed by Narayan et al. (1997), except that they included an
additional differential equation for $a(r)$, based on the entropy
equation~(\ref{eq13}). However, this extra differential equation is not
necessary because the energy flow rate $\enrate$ is conserved when
radiative losses are neglected, as assumed in the ADAF scenario
(Molteni, Gerardi, \& Valenza 2001). This fact allows us to solve for
$a$ as an algebraic function of $\vr$, $\ell$, and $r$ using
equation~(\ref{eq41}), which yields
\begeq
a^2 = {\gamma-1 \over \gamma} \left(\epsilon_0 - {\vr^2 \over 2}
+ {\ell^2 \over 2 \, r^2} - {\ell_0 \ell \over \, r^2}
+ {GM \over r-\rs}\right) \ .
\label{eq51}
\fineq

Critical points occur where the numerator and denominator in equation~(\ref{eq49})
for $d\ln v/dr$ vanish simultaneously. This yields the critical conditions
\begeq
{\vr^2 \over a^2} - {2 \, \gamma \over \gamma+1} = 0 \ , \ \ \ \ \ 
r = \rc \ ,
\label{eq52}
\fineq
\begeq
{\ell^2 - \ellK^2 \over a^2 \, r^3}
+ {2 \, \gamma \over \gamma+1} \left({3 \over r} - {d\ln\ellK
\over dr}\right) - \left(\gamma-1 \over \gamma+1\right) {\vr \,
\ellK \, (\ell - \ell_0)^2 \over \alpha \, a^4 \, r^4} = 0 \ ,
\ \ \ \ \ r = \rc \ ,
\label{eq53}
\fineq
where $r_c$ is the critical radius. Global solutions must pass through a
critical point, and therefore equations~(\ref{eq51}), (\ref{eq52}), and
(\ref{eq53}) can be used to interrelate the six quantities $(r_c, \vc ,
a_c, \ell_c, \epsilon_0, \ell_0)$, where $\vc$, $a_c$, and $\ell_c$
denote quantities measured at the critical radius $r=r_c$. We can
integrate the system of equations~(\ref{eq27}) and (\ref{eq49}) away
from the critical point, either towards large radii or towards the
horizon. However, due to the nature of the critical point, we cannot
begin the integration precisely at $r=\rc$. We must therefore employ
l'H\^opital's rule to evaluate $d\vr/dr$ at the critical point, and then
perform a linear extrapolation to offset the starting conditions
slightly in radius (Molteni et al. 2001; Chen \& Taam 1993; Chen et al.
1997). This procedure involves the solution of a quadratic equation for
the critical value of $dv/dr$. In our application, the negative, real
root gives the value for the derivative at the critical point.

\subsection{Disk Structure Calculations}

In order to illustrate the utility of the asymptotic relations developed
in \S~4, we shall perform several calculations of the disk structure
based upon explicit integration of the differential equations. Since
there are two coupled differential equations in the system, there are
two linearly independent local solutions around the singular point at
the horizon. Only one of the local solutions is physically acceptable,
and this is the solution that we have obtained asymptotic representations
for in \S~4. Explicit integration of the system of equations from the
critical point towards the horizon is unstable, because in general it is
impossible to avoid exciting the second, linearly independent solution,
which displays an unphysical behavior as the gas approaches the horizon.
While it is an open question whether shocks supported by a ``centrifugal
barrier'' exist in black hole accretion disks (e.g., Chakrabarti 1997),
our goal here will be to develop global, shock-free solutions in order
to demonstrate the utility of the boundary conditions derived in \S~4
in the simplest possible manner.

With the availability of the asymptotic expressions for $\vr(r)$ and
$\ell(r)$ derived in \S~4, we can employ an explicit, stable integration
in the {\it outward} direction, starting at a point just outside $r =
\rs$. In this approach, equations~(\ref{eq35}) and (\ref{eq44}) are used
to set the starting values for $\ell$ and $\vr$, respectively, as
functions of the three constants $(\epsilon_0, \ell_0, K_0)$. Numerical
integration of equations~(\ref{eq27}), (\ref{eq49}), and (\ref{eq51}) in
the outward direction yields solutions for $\ell(r)$, $v(r)$, and
$a(r)$. For given values of $\ell_0$ and $\epsilon_0$, the parameter
$K_0$ can be determined by requiring that the flow pass smoothly through
a critical point, at some radius $r=r_c$. In order to determine $\ell_0$
and $\epsilon_0$, we must therefore supply two additional boundary
conditions. These extra conditions are usually imposed by requiring that
the disk become Keplerian and geometrically thin at some arbitrary outer
radius (Narayan et al. 1997). However, it is not completely clear
whether ADAF solutions can merge smoothly with cool thin disks (Yuan
1999; Yuan et al. 2000). This particular issue is not central to our
considerations in this paper, since our focus here is on discussing the
inner boundary conditions appropriate for advection-dominated black hole
accretion. Therefore, in order to develop numerical examples that
illustrate the role of the inner boundary condition without undue
complexity, we shall simply set $\epsilon_0 = 0$ and require that $a \to
0$ as $r \to \infty$, in keeping with the self-similar,
advection-dominated models (Narayan \& Yi 1994, 1995; Blandford \&
Begelman 1999; Becker, Subramanian, \& Kazanas 2001). The
self-consistency of the numerical solutions obtained for $\ell(r)$,
$v(r)$, and $a(r)$ using the outward integration is checked by
confirming that at the critical radius, $\ell(r_c)=\ell_c$, $v(r_c) =
\vc$, and $a(r_c) = a_c$, where the quantities $(\vc, a_c, \ell_c, r_c)$
satisfy the critical conditions given by equations~(\ref{eq51}),
(\ref{eq52}), and (\ref{eq53}).

The integrations begin at a starting radius, $r_*$, located close to the
event horizon, and proceed in the outward direction, towards the
critical point. In our numerical examples, we shall work in terms of
natural gravitational units ($GM=c=1$, $\rs=2$), and the starting radius
will be given by $r_* = 2.001$, which is just outside the Schwarzschild
radius. The corresponding starting values for the specific angular
momentum $\ell_* = \ell(r_*)$ and the inflow velocity $v_* = v(r_*)$ are
computed by applying the asymptotic formulas given by
equation~(\ref{eq35}) and (\ref{eq44}), respectively, at the radius
$r=r_*$. The starting value for the isothermal sound speed, $a_* =
a(r_*)$, is determined using equation~(\ref{eq51}). The exact solutions
for $\vr$, $a$, and $\ell$ as functions of radius are obtained by
integrating numerically the coupled equations~(\ref{eq27}),
(\ref{eq49}), and (\ref{eq51}). The values of the various model
parameters are listed in Table~2 for the three models that we consider
in detail below. The specific parameter values we have selected
correspond to those adopted by Narayan et al. (1997) in several of their
calculations.

In Figures~1, 2, 3, 4, and 5, we present solutions obtained by setting
$\ell_0=2.6$, $\epsilon_0=0$, $\alpha=0.1$, and $\gamma=1.5$, which we
refer to as Model~1. Note that this value for $\gamma$ represents
approximate equipartition between the gas pressure and the magnetic
pressure. In this scenario, the $\alpha$-prescription we have employed
for the viscosity (eq.~[\ref{eq8}]) is valid even within the supersonic
region of the flow. The exact numerical solution for the inflow velocity
$v(r)$ is compared with the asymptotic solution given by
equation~(\ref{eq44}) in Figure~1. Also included for comparison are the
free-fall velocity distribution, $\vff(r) = [2GM/(r-\rs)]^{1/2}$, and
the exact numerical solution for the isothermal sound speed $a(r)$. The
agreement between the exact solution for $v(r)$ and the asymptotic
expression is excellent for $2 < r < 3$, and remains reasonably close all
the way out to the critical point, which is located at $r_c = 6.132$ for
this model. Note that $v$ remains well below $\vff$ until the material
gets quite close to the horizon. The exact numerical solution for the
sound speed $a(r)$ is compared with the asymptotic solution
(eq.~[\ref{eq47}]) in Figure~2. The agreement between these two results
is surprisingly close all the way out to the critical radius. In
Figure~3 we plot the exact numerical solution for the specific angular
momentum, $\ell = r^2 \, \Omega$, along with the asymptotic formula
given by equation~(\ref{eq35}). The two expressions for $\ell(r)$ merge
smoothly as $r \to \rs$, in validation of our asymptotic analysis. In
Figure~4 we display the global solutions obtained by joining the
numerical results for $v(r)$ and $a(r)$ from Figure~1 with curves
generated by integrating in the outward direction starting from the
critical radius $r = r_c$. Note that the global solutions pass smoothly
through the critical point as required. In Figure~5 we display the
results obtained for $v(r)$ and $a(r)$ by integrating in the {\it
inward} direction starting from the critical point. These are plotted
along with the numerical solutions obtained using the outwardly-directed
integration starting close to the horizon (Fig.~1). Near the critical
point, the two sets of solutions agree closely. However, the
inwardly-directed integration is unstable, and the denominator of the
dynamical equation~(\ref{eq49}) vanishes at $r \sim 5$, where $v^2 = 2
\, \gamma \, a^2 / (1+\gamma)$. The existence of this instability
provides one of the main motivations for developing the inner boundary
conditions in \S~4, since the availability of these boundary conditions
facilitates the integration in the outward direction, which is stable.

In Figures~6, 7, 8, and 9, we present results for Model~2, with
$\ell_0=1.76$, $\epsilon_0=0$, $\alpha=0.3$, and $\gamma=1.5$. The exact
and asymptotic solutions for the inflow velocity $v(r)$ are compared in
Figure~6. Note that two results are indistinguishable in the entire
region between the horizon and the sonic (critical) point, which is
located at $r_c = 10.63$ for the Model~2 parameters. The exact solution
for the sound speed $a(r)$ obtained via numerical integration is
compared with the asymptotic expression (eq.~[\ref{eq47}]) in Figure~7.
In this case, the two results agree closely for $2 < r < 3$, although
the agreement deteriorates for $r \gapprox 5$. The exact numerical
solution for the specific angular momentum $\ell = r^2 \, \Omega$ is
plotted in Figure~8 along with the asymptotic result given by
equation~(\ref{eq35}). The two expressions again display a smooth merger
as the gas approaches the horizon. In Figure~9 we plot complete global
solutions obtained by combining the Figure~6 results with solutions for
$v(r)$ and $a(r)$ obtained by integrating away from the critical point
in the outward direction. The global solutions pass smoothly through the
critical point, in satisfaction of the critical conditions.

It is interesting to compare our results with those obtained by Narayan
et al. (1997). An examination of their Figures~1, 2, and 3 reveals close
agreement with our results. In fact, their plots of the variation of
$\ell$ near the horizon all show that $d\ell/dr \to 0$ as $r \to \rs$,
which is consistent with our prediction based on equation~(\ref{eq35}).
This is intriguing considering the fact that they did not formally adopt
boundary conditions identical to ours in their calculations. In fact,
they state that $d\Omega/dr = 0$ at the horizon in the discussion
following their equation~(2.18), which is obviously incorrect. Hence
their treatment of the inner boundary conditions is unclear. Our
numerical results also agree with those of Chen et al. (1997), although
these authors do not extend their calculations to the horizon, and
instead truncate the disk at an arbitrary radius $r_{\rm in} = 3$. For
the starting radius in our outwardly-directed integrations, we have set
$r_* = 2.001$, and the corresponding local free-fall velocity there is
$\vff(r_*) = 44.7214$. The values we obtain for the starting velocity
$v_*$ displayed in Table~2 are just slightly smaller than the local
free-fall velocity, as a result of centripetal and pressure effects.
However, this small difference is crucial for determining the velocity
distribution via the subsequent integration away from the horizon.

Along with the results associated with Models~1 and 2 results, Table~2 also
includes a summary of the results obtained for Model 3, with $\ell_0=3.21$,
$\epsilon_0=0$, $\alpha=0.03$, and $\gamma=1.5$. It is interesting to compare
the values obtained for the entropy and angular momentum parameters in the
three models. In particular, Table~2 includes results for the entropy function
$K(r)$ (eq.~[\ref{eq14}]) evaluated at the critical radius, $K_c \equiv K(r_c)$,
as well as at the horizon, $K_0 \equiv K(\rs)$. Note that $K_0$ always exceeds
$K_c$, reflecting the fact that viscous dissipation between the sonic point and
the horizon increases the entropy of the gas. Our results indicate that $K_0$
exceeds $K_c$ by 91\% for $\alpha = 0.3$, by 37\% for $\alpha = 0.1$, and by 14\%
for $\alpha = 0.03$. This is consistent with the expectation that larger values
of $\alpha$ should be associated with higher dissipation rates. Also note that
the accreted specific angular momentum, $\ell_0$, is always lower than the
specific angular momentum at the sonic point, $\ell_c$, and that the ratio
$\ell_0 / \ell_c$ decreases with increasing $\alpha$, as expected, due to
the increase in the magnitude of the viscous torque.

\section{DISCUSSION}

The development of global solutions for the structure of
advection-dominated accretion disks requires careful consideration of
the boundary conditions that apply close to the black hole event
horizon, because the horizon always represents the fundamental inner
boundary for any such calculation. In this paper we have explored the
consequences of the pseudo-Newtonian potential for the asymptotic
structure (in the vicinity of the horizon) of advection-dominated
accretion disks incorporating the Shakura-Sunyaev viscosity
prescription. The presence of the pseudo-Newtonian potential introduces
a regular singular point into the differential conservation equations.
In recognition of this fact, we have employed asymptotic analysis to
determine the leading behaviors of the physical quantities $\vr$, $a$,
$\ell$, and $K$ close to the horizon. The main asymptotic results
derived in \S~4 are given by equation~(\ref{eq35}) for $\ell(r)$,
equation~(\ref{eq40}) for $K(r)$, equation~(\ref{eq44}) for $v(r)$, and
equation~(\ref{eq47}) for $a(r)$. These expressions clearly illustrate
how the various physical quantities depend on the three parameters
$\epsilon_0$, $\ell_0$, and $K_0$, denoting the accreted specific
energy, the accreted specific angular momentum, and the accreted
specific entropy, respectively.

The analytical, asymptotic relations we have obtained can be used to
provide the inner boundary conditions needed for simulation of the disk
structure. These boundary conditions are essential whether the
computation is performed using an explicit integration or an iterative
relaxation technique. In \S~5.2, the new boundary conditions were used
to perform an explicit numerical integration of the coupled conservation
equations in the outward direction, starting at a point just outside the
event horizon and ending at the critical point. The exact numerical
solutions obtained for $\vr$, $a$, and $\ell$ were compared with the
corresponding analytical expressions derived in \S~4. The agreement
between the two sets of solutions is remarkable, providing positive
confirmation of the validity of our asymptotic analysis and the
resulting boundary conditions. We have demonstrated in Appendix~A that
the velocity distribution associated with our pseudo-Newtonian model
agrees with general relativity in the vicinity of the horizon. In the
remainder of this section, we will examine a few additional questions
related to the self-consistency of the model.

\subsection{Torque Boundary Condition}

In our model, the torque ${\cal G}$ vanishes at the horizon, rather than
at the radius of marginal stability, $r_{\rm ms} = 6 \,GM/c^2$. This is
consistent with simulations performed by Hawley \& Krolik (2001), Agol
\& Krolik (2000), and Gammie (1999), who all find that the stress has a
finite value at $r = r_{\rm ms}$. Furthermore, our results demonstrate
that any advection-dominated disk with hydrostatic vertical structure
must have $d\ell/dr=0$ at the horizon. This in turn implies that the
radial derivative of the torque must vanish there, i.e., $d{\cal
G}/dr=0$ at $r=\rs$ (eq.~[\ref{eq38}]). This new condition supplements
the well-known requirement that ${\cal G}=0$ on the horizon. We show
below that this behavior can be understood as a simple consequence of
the adiabatic nature of the flow close to the horizon. First we combine
the fundamental expression for the torque, equation~(\ref{eq7}), with
the mass conservation equation~(\ref{eq2}) and the Shakura-Sunyaev
viscosity prescription (eq.~[\ref{eq8}]) to obtain
\begeq
{\cal G} = - {\alpha \, \dot M \, a^2 \, r^2 \over v \, \OmegaK}
{d\Omega \over dr} \ .
\label{eq54}
\fineq
We have found that close to the horizon, the viscous heating vanishes
and therefore the pressure obeys the adiabatic relation $P \propto
\rho^\gamma$, implying that $a^2 \propto \rho^{\gamma-1}$. Substituting
for $\rho$ using the mass conservation relation $\dot M = 4 \pi r H \rho
v$, we find that $a^2 \propto (rHv)^{1-\gamma}$ as $r \to \rs$. In a
hydrostatic disk, $H \propto a / \OmegaK$, and consequently $a \propto
(r \, v / \OmegaK)^{(1-\gamma)/(1+\gamma)}$ (cf. eq.~[\ref{eq42}]).
Since $d\Omega/dr$ approaches a finite value as $r \to \rs$, and $\vr$
approaches free-fall, we immediately conclude that ${\cal G} \propto
(r-\rs)^\beta$, with $\beta = (\gamma+5)/(2+2\gamma)$. This agrees with
our earlier derivation (see Table~1), and confirms that $\beta > 1$, with
the resulting implication that $d{\cal G}/dr$ must vanish at the horizon.
Hence we have demonstrated that the vanishing of the derivative of the
torque at the horizon is a predictable consequence of the scalings of
$a$ and $\vr$ as $r \to \rs$ in standard ADAF disks.

\subsection{Effects of Central Free-Fall}

In this paper we have focused on the behavior of advection-dominated
disks that are in vertical hydrostatic equilibrium at all radii, because
we are interested in developing inner boundary conditions that are
consistent with the standard ADAF scenario (Narayan et al. 1997; Chen et
al. 1997). The boundary conditions and asymptotic behaviors we have
obtained agree well with global ADAF simulations, and therefore they
provide a useful foundation for numerical calculations of the disk
structure. While our approach is a reasonable strategy from a
computational point of view, we should acknowledge that in reality, the
gas will probably stop responding to vertical pressure forces in the
supersonic region close to the horizon. It is perhaps more likely that
the gas crosses the horizon in nearly radial free-fall, with $H \propto
r$. From a physical point of view, it is worthwhile to consider how this
alternative central inflow condition would affect some of the basic
conclusions we have reached in this paper.

We have shown quite generally in \S~4.3 that the dissipation rate vanishes
as the gas approaches the horizon, and therefore $P \propto \rho^\gamma$
and $a^2 \propto \rho^{\gamma-1}$. This result does not depend on the
assumption of vertical hydrostatic equilibrium. Proceeding as in \S~6.1,
we substitute for $\rho$ using the mass conservation equation~(\ref{eq2})
and find that, in general, $a^2 \propto (r H v)^{1-\gamma}$ as $r \to \rs$.
In the quasi-spherical free-fall region close to the horizon, the disk
half-thickness is given approximately by
\begeq
H = d_0 \, r \ ,
\label{eq59.1}
\fineq
where $d_0$ is a constant. This implies that as $r \to \rs$, the variation
of the sound speed satisfies
\begeq
a^2 \propto (r^2 \, v)^{1-\gamma} \ .
\label{eq59.2}
\fineq
In the free-fall case, it is convenient to define the entropy function using
the alternative form
\begeq
\tilde K(r) \equiv r^2 \, v \, a^{2/(\gamma-1)} \ .
\label{eq59.3}
\fineq
Although differs from the definition of $K(r)$ used in the hydrostatic scenario
(eq.~[\ref{eq14}]), by combining equations~(\ref{eq2}), (\ref{eq9}), (\ref{eq59.1}),
and (\ref{eq59.3}), we can confirm that $\tilde K^{\gamma-1} \propto U
\rho^{-\gamma}$, and therefore $\tilde K$ is a linear function of the
entropy per particle $S$ if the disk is in free-fall (see eqs.~[\ref{eq15}]
and [\ref{eq17}]). Next we express the asymptotic behavior of $\ell(r)$ close
to the horizon using
\begeq
\ell(r) \asym \ell_0 + \tilde B (r-\rs)^{\tilde \beta} \ ,
\label{eq59.4}
\fineq
where $\ell_0$ is the accreted specific angular momentum, and the constants
$\tilde B$ and $\tilde \beta$ are analogous to the hydrostatic constants
$B$ and $\beta$ appearing in equation~(\ref{eq28}). Following the same steps
as in \S~4.3, we now obtain
\begeq
\lim_{r \to \rs} {\vr \, \OmegaK \over \alpha \, a^2} \, \tilde B \,
(r-\rs)^{\tilde \beta} = {2 \, \ell_0 \over \rs} - \ellprime0 \ .
\label{eq59.5}
\fineq
Using asymptotic analysis and requiring that exponents and coefficients
balance on the two sides of this expression, it is straightforward to show
that the solutions for $\tilde B$ and $\tilde \beta$ are given by
\begeq
\tilde \beta = {\gamma + 2 \over 2} \ , \ \ \ \ \ \ 
\tilde B = {8^{1/2} \, \alpha \, \ell_0 \, \rs \over c \, \tilde K_0}
\left(\tilde K_0 \over c \, \rs^{5/2} \right)^\gamma \ ,
\label{eq59.6}
\fineq
where $c$ is the speed of light and
\begeq
\tilde K_0 \equiv \lim_{r \to \rs} \tilde K(r)
\label{eq59.7}
\fineq
denotes the value of the entropy function at the horizon. Note that
$\tilde\beta > 1$ for all $4/3 \le \gamma \le 5/3$, and consequently
$d{\cal G}/dr= d\ell/dr=0$ at the horizon. This is the same conclusion
reached in \S~4.3 under the assumption of vertical hydrostatic equilibrium,
although the exponent $\tilde\beta$ differs slightly from the hydrostatic
exponent $\beta=(\gamma+5)/(2+2\gamma)$. Hence the vanishing of the derivative
of the torque at the horizon in ADAF disks is a very general result.

\subsection{Causality of Viscous Transport at the Horizon}

Numerous authors have pointed out that the diffusive nature of the
angular momentum transport associated with the Shakura-Sunyaev viscosity
prescription $\nu = \alpha \, a^2/\OmegaK$ can lead to causality
violations in accretion disks (e.g., Kato 1994; Narayan 1992). This
issue can be most easily understood by considering the evolution of an
initially localized component of the angular momentum distribution,
represented by a $\delta$-function at some arbitrary radius $r=r_0$ and
arbitrary time $t=t_0$. As time proceeds, the distribution will spread
in radius in an approximately Gaussian manner, implying propagation to
infinite distance in a finite time, which violates causality. This
phenomenon has a negligible effect on the structure of the disk in the
outer, subsonic region, because the {\it mean} transport velocity for
the angular momentum is typically very small despite the fact that an
infinitesimal portion of the signal propagates with infinite speed.
However, the question of causality needs to be examined carefully in the
inner, supersonic region, where the radial inflow velocity $v^{\hat r}$
approaches the speed of light, and all signals should therefore be
advected into the black hole. In order to address this issue in the
context of the ADAF scenario considered here, it is useful to examine
the diffusion equation governing the angular momentum distribution in
the disk. Following Blandford \& Begelman (1999), we write the
time-dependent equation as
\begeq
{\partial \mu r^2 \Omega \over \partial t} = {\partial \over \partial r}
\left(\mu r^2 \Omega v - {\cal G} \right) \ ,
\label{eq55}
\fineq
where
\begeq
\mu \equiv 4 \pi r \rho = {\dot M \over v}
\label{eq55.1}
\fineq
represents the mass per unit radius in the disk. By combining equation~(\ref{eq55})
with equation~(\ref{eq7}) for the torque ${\cal G}$, we can obtain the alternative
form
\begeq
{\partial L \over \partial t} = {\partial \over \partial r}
\left[v \, L + \nu \, {\partial L \over \partial r} - {\nu L \over \mu}
{\partial \mu \over \partial r} - {2 \nu L \over r} \right] \ .
\label{eq55.7}
\fineq
where
\begeq
L \equiv \mu r^2 \Omega
\label{eq55.5}
\fineq
denotes the angular momentum per unit radius.

To obtain further insight, we can recast equation~(\ref{eq55.7}) in the form
of the Fokker-Planck equation
\begeq
{\partial L \over \partial t} = - {\partial \over \partial r}
\left({d \langle r \rangle \over dt} L \right) + {\partial^2 \over \partial r^2}
\left({1 \over 2}{d \sigma^2 \over dt} L \right) \ ,
\label{eq55.71}
\fineq
where the Fokker-Planck coefficients
\begeq
{d \langle r \rangle \over dt} = {2 \nu \over r} + {\partial \nu \over
\partial r} + {\nu \over \mu} {\partial \mu \over \partial r} - v \ ,
\ \ \ \ \ \ \ \ 
{1 \over 2}{d \sigma^2 \over dt} = \nu \ ,
\label{eq55.73}
\fineq
describe, respectively, the rates of ``drifting'' and ``broadening''
experienced by the initially localized angular momentum distribution due
to diffusion (Reif 1965). We are interested in evaluating the
Fokker-Planck coefficients in the context of the steady-state ADAF disks
considered here, which have $\mu \propto v^{-1}$. In the inner region,
close to the horizon, we have found that $v \propto \vff =
[2GM/(r-\rs)]^{1/2}$ and $a \propto (r-\rs)^{(1-\gamma)/(2\gamma+2)}$.
Consequently, the Shakura-Sunyaev viscosity, $\nu = \alpha \,
a^2/\OmegaK$, displays the asymptotic behavior $\nu \propto
(r-\rs)^{2/(1+\gamma)}$ in the vicinity of the horizon. Since $\nu$
vanishes as $r \to \rs$, it follows that $d \sigma^2/dt \to 0$. One can
also show that the mean transport velocity, $d\langle r \rangle/dt$,
approaches $-v$ as $r \to \rs$. The vanishing of the ``broadening'' rate
at the horizon implies that the angular momentum is simply advected into
the black hole, and there is no nonphysical transport to infinite
distance in finite time. Taken together, our asymptotic results for the
Fokker-Planck coefficients $d \sigma^2/dt$ and $d\langle r \rangle/dt$
demonstrate that the transport of angular momentum at the horizon is
{\it causal}, in agreement with general relativity. We have therefore
confirmed that, at least in the context of the ADAF model considered
here, there are no causality violations at the horizon associated with
the Shakura-Sunyaev prescription for the viscosity. Interestingly, this
result remains valid even if the hydrostatic relation is replaced with
the central free-fall condition discussed in \S~6.2.

\subsection{Energy and Mass Loss}

In this paper, we have adopted the perfect ADAF approximation by
assuming that all of the transfer rates $\enrate$, $\momrate$, and $\dot
M$ are constant. We have therefore neglected the possibility of
radiative losses, as well as outflows of matter and energy. This
contradicts recent observations suggesting that many low-luminosity
X-ray AGNs possess relativistic jets that originate very close to the
event horizon of the central black hole (Fender 2001; Nagar et al.
2002). Examples include RS 1915+105 (Belloni et al. 1997; Dhawan,
Mirabel, \& Rodr\'iguez 2000), XTE J1118+480 (Fender et al. 2001), XTE
J1550-564 (Corbel et al. 2001), GS 1354-64 (Brocksopp et al. 2001), and
perhaps Cyg X-1 (Stirling et al. 2001). The fact that disks and jets
often appear together suggests that the presence of the outflows may be
a necessary ingredient for the accretion to proceed. This possibility
has motivated several investigations into the relationship between
outflows and ADAFs. From a theoretical viewpoint, the positivity of the
Bernoulli parameter in advection-dominated flows suggests that the gas
in these systems is gravitationally unbound (Narayan, Kato, \& Honma
1997; Narayan \& Yi 1994, 1995). Based on this observation, Blandford \&
Begelman (1999) investigated the effects of mass, energy, and angular
momentum loss on the structure of the disk in the context of a
self-similar model incorporating Newtonian gravity. This approach was
extended by Becker et al. (2001) to describe self-similar disk/outflow
systems governed by the pseudo-Newtonian potential.

Observations of strong radio emission from X-ray underluminous AGNs
suggest that relativistic particles abound in the hot plasmas. The
typical energy of these particles is much higher than the average
thermal energy of the gas, implying the presence of an efficient
acceleration mechanism. In this connection, it is interesting to note
that the low density in advection-dominated disks makes them plausible
sites for the acceleration of relativistic particles via interactions
with magnetohydrodynamical waves because the plasma is so tenuous that
ion-ion collisions are unable to thermalize the energy of the
accelerated particles (Becker et al. 2001; Subramanian,
Becker, \& Kazanas 1999). Hence in the X-ray underluminous AGNs, particle
acceleration and the resulting outflows of unbound particles from the
disk may represent the dominant cooling mechanism, removing excess
energy and thereby allowing accretion to proceed. Conversely, in the
X-ray bright systems, the efficiency of particle acceleration in the
disk is lower due to the higher density, which tends to thermalize the
energy of the accelerated particles. In these systems, it is the X-ray
emission that removes most of the binding energy. This interpretation
helps to explain the observed anticorrelation between the outflow
strength and the X-ray luminosity, as well as the positive correlation
between the X-ray hardness ratio and the radio emission strength
(Celotti \& Blandford 2001; Corbel et al. 2000).

While outflows have not been incorporated into our analysis, we expect
that their inclusion would have little if any effect on the boundary
conditions at the event horizon that we have derived. This is because
the power source for the outflows would presumably be the viscous
dissipation, which clearly vanishes rather quickly below $r=r_{\rm ms}$.
Hence the physics in the asymptotic region close to the horizon should
be insensitive to the production of the outflows. In future work, we
plan to utilize our asymptotic relations to facilitate the development
of detailed disk models including outflows of matter and energy that are
self-consistently coupled with the disk.

\subsection{Conclusion}

In this paper we have obtained a number of useful analytical expressions
that completely describe the structure (close to the event horizon) of
advection-dominated, pseudo-Newtonian accretion flows based on the
Shakura-Sunyaev viscosity prescription. The dynamical results depend
on three quantities: the accreted specific energy $\epsilon_0$, the
accreted specific angular momentum $\ell_0$, and the accreted specific
entropy $k \ln K_0$. In our approach, $\epsilon_0$ and $\ell_0$ are
treated as free parameters, and the value of $K_0$ is determined by
requiring that the global solution pass smoothly through a critical
point. The asymptotic expressions derived in \S~4 provide a set of inner
boundary conditions that can serve as the basis for subsequent numerical
integration of the conservation equations. We emphasize that any
physically consistent one-dimensional advection-dominated accretion disk
model based on the Shakura-Sunyaev viscosity prescription {\it must}
satisfy these boundary conditions. The analytical expressions agree
extremely well with the exact numerical solutions out to a few
gravitational radii from the horizon. Hence our results provide a valid
description of the essential physics of the accretion process in the
vicinity of the horizon.

We have assumed that the effects of general relativity can be
approximated using the pseudo-Newtonian gravitational potential. While
this is a widely used approximation that preserves many of the important
dynamical characteristics of flows in the Schwarzschild metric, one may
well ask how the specific results we have obtained here translate into
full general relativity. For example, will the radial derivative of the
torque really vanish at the horizon in the Schwarzschild metric?
Obviously this question cannot be answered definitively without
employing a fully relativistic calculation. However, we demonstrate in
Appendix~A that the motions of particles near the horizon predicted by
our pseudo-Newtonian model are in complete agreement with the actual
motions of freely-falling particles in the Schwarzschild metric.
Furthermore, we have established in \S~6.3 that the diffusive transport
of angular momentum in our model is causal in nature at the horizon.
This confirms that pseudo-Newtonian ADAF models incorporating the
Shakura-Sunyaev viscosity prescription can be used to describe the
structure of the disk all the way to the event horizon. We argue based
on these results that the general characteristics of the asymptotic
solutions we have obtained are likely to be preserved in full general
relativity.

The numerical examples presented in \S~5 do not include shocks, which
may occur in accretion flows as a result of the interaction between the
gas and a ``centrifugal barrier'' (Chakrabarti 1997). Shocks may also
play a role in powering the outflows associated observationally with hot
disks (Yuan et al. 2002). Although shocks were not explicitly considered
in our analysis, we argue that our asymptotic results should apply
equally well whether or not shocks are present, because our results are
based on the fundamental physical processes operative near the horizon,
and those processes are insensitive to the history of the gas. A related
question concerns the relevance of our results in the context of
convection-dominated accretion flows (``CDAFs''). These are stationary,
convective-envelope solutions that technically have zero accretion
rates, although in fact a small amount of matter is expected to flow
into the black hole (Narayan, Igumenshchev, \& Abramowicz 2000). We
emphasize that our basic results should apply in this situation as well
because the gas that enters the black hole must satisfy the same
asymptotic conservation equations, independent of whether it has passed
through a CDAF, a shocked disk, or a conventional ADAF.

The authors are grateful to the anonymous referee for helping us
to improve the discussion. PAB would also like to acknowledge several
stimulating conversations with Menas Kafatos, Ken Wolfram, and Demos
Kazanas, as well as generous support from the Naval Research Laboratory
during a portion of the research.

\eject

\appendix

\section{APPENDIX}

\centerline{PSEUDO-NEWTONIAN PARTICLE DYNAMICS}

In our approach to modeling the disk structure, we have incorporated
the effects of general relativity in an approximate manner by utilizing
the pseudo-Newtonian potential,
\begeq
\Phi(r) = {- G \, M \over r - \rs} \ ,
\label{eq1.0}
\fineq
which gives correct results for the radius of marginal stability
($r_{\rm ms} = 6 \, GM/c^2$), the marginally bound radius ($r_{\rm mb} =
4 \, GM/c^2$), and the horizon radius ($\rs = 2 \, GM/c^2$) around a
nonrotating black hole (Paczy\'nski \& Wiita 1980; Abramowicz, Calvani,
\& Nobili 1980). In order to understand how the motions of particles in
the pseudo-Newtonian potential are related to the exact solutions given
by general relativity, we shall briefly review the dynamics of particles
freely falling in the Schwarzschild metric. We begin by writing down
exact expressions for the radial velocity $v^{\hat r}$ and the azimuthal
velocity $v^{\hat \varphi}$ describing the motion of a particle as
measured by a local, static observer in the Schwarzschild geometry.
Using equations~(12.4.17) and (12.4.18) from Shapiro \& Teukolsky
(1983), we obtain
\begeq
v^{\hat r} = c \left[1 - \left(E \over c^2\right)^{-2}
\left(1 - {\rs \over r}
\right)\left(1 + {\ell^2 \over c^2 r^2}\right) \right]^{1/2} \ ,
\label{eq1.1}
\fineq
and
\begeq
v^{\hat \varphi} = c \left(1 - {\rs \over r}\right)^{1/2}
{\ell \, c \over r E} \ ,
\label{eq1.2}
\fineq
where $E$ and $\ell$ denote the particle's specific energy and specific
angular momentum at infinity, respectively. The locally measured
specific energy, $E_{\rm local}$, is related to $E$ by
\begeq
E_{\rm local} = E \left(1 - {\rs \over r}\right)^{-1/2} \ .
\label{eq1.3}
\fineq
In terms of the locally measured Lorentz factor,
\begeq
\Gamma_{\rm local} \equiv {E_{\rm local} \over c^2} \ ,
\label{eq1.3a}
\fineq
the radial component of the {\it four-velocity}, $v_r$, measured by
a static observer at radius $r$ can be written as
\begeq
v_r \equiv \Gamma_{\rm local} \, v^{\hat r}
= {E \over c} \left(1 - {\rs \over r}\right)^{-1/2}
\left[1 - \left(E \over c^2\right)^{-2} \left(1 - {\rs \over r}\right)
\left(1 + {\ell \over c^2 r^2}
\right) \right]^{1/2} \ ,
\label{eq1.4}
\fineq
and the locally measured azimuthal component of the four-velocity,
$v_\varphi$, is given by
\begeq
v_\varphi \equiv \Gamma_{\rm local} \, v^{\hat \varphi}
= {\ell \over r} \ .
\label{eq1.5}
\fineq

We shall now focus on the case of a particle falling from rest at
infinity, with $E = c^2$. Note that $\ell$ can still have an arbitrary
value in this case, since the azimuthal velocity vanishes at infinity
for all values of $\ell$. Our expression for $v_r$ now reduces to
\begeq
v_r^2 = {2 \, G M \over r - \rs} - {\ell^2 \over r^2} \ ,
\label{eq1.6}
\fineq
or, equivalently,
\begeq
{1 \over 2} \, v_r^2 + {1 \over 2} \, v_\varphi^2 + \Phi(r) = 0 \ ,
\label{eq1.7}
\fineq
where $\Phi(r)$ is the pseudo-Newtonian potential given by
equation~(\ref{eq1.0}). Equation~(\ref{eq1.7}) resembles
a classical Newtonian energy equation, except that $v_r$ and
$v_\varphi$ are four-velocities rather than conventional velocities.
This is, in fact, one of the basic motivations for introducing the
pseudo-Newtonian potential. Despite its classical appearance, the
result obtained for $v_r$ by solving equation~(\ref{eq1.7}) with
$v_\varphi = \ell/r$ is exactly equal to the radial four-velocity
of a particle freely falling from rest at infinity in the Schwarzschild
metric. Note that as the particle approaches the event horizon
of the black hole, the radial four-velocity diverges, i.e.,
\begeq
v_r(r) \to \vff(r) \equiv \left(2 \, GM \over r - \rs \right)^{1/2} \ ,
\ \ \ \ \ \ \ 
r \to \rs \ .
\label{eq1.8}
\fineq
On the other hand, the azimuthal four-velocity, $v_\varphi$, remains
bounded, and we find that for a given value of $\ell$,
\begeq
v_\varphi(r) \to {\ell \over \rs} \ ,
\ \ \ \ \ \ \ 
r \to \rs \ .
\label{eq1.9}
\fineq
As the particle approaches the horizon, the limiting values for the
physical velocities $v^{\hat r}$ and $v^{\hat \varphi}$ are given by
\begeq
\lim_{r \to \rs} v^{\hat r} = c \ , \ \ \ \ \ \ 
\lim_{r \to \rs} v^{\hat \varphi} = 0 \ ,
\label{eq1.10}
\fineq
which follow from equations~(\ref{eq1.1}) and (\ref{eq1.2}). Hence
a stationary observer close to the horizon sees the particle falling
radially inward at the speed of light, as expected.

As a consequence of utilizing the pseudo-Newtonian potential to
describe ADAF disks, we have found in \S~4 that the asymptotic
behavior of the ``radial velocity'' $v$ is given by
\begeq
v \to \left(2 \, GM \over r-\rs\right)^{1/2} \ ,
\ \ \ \ \ \ \ r \to \rs \ ,
\label{appen1}
\fineq
and the asymptotic behavior of the ``azimuthal velocity'' $w = r \, \Omega$
is likewise given by
\begeq
w \to {\ell_0 \over \rs} \ , \ \ \ \ \ r \to \rs \ ,
\label{appen2}
\fineq
where $\ell_0 = \dot J / \dot M$ is the specific angular momentum of the
material crossing the event horizon. Comparing these expressions with
equations~({\ref{eq1.8}) and (\ref{eq1.9}), we find that close to the
horizon, the quantities $v$ and $w$, respectively, are exactly equal to
the radial component ($v_r$) and the azimuthal component ($v_\varphi$)
of the four-velocity for a particle freely falling in the
Schwarzschild metric.

\clearpage

\epsscale{0.8}

\begin{figure}
\plotone{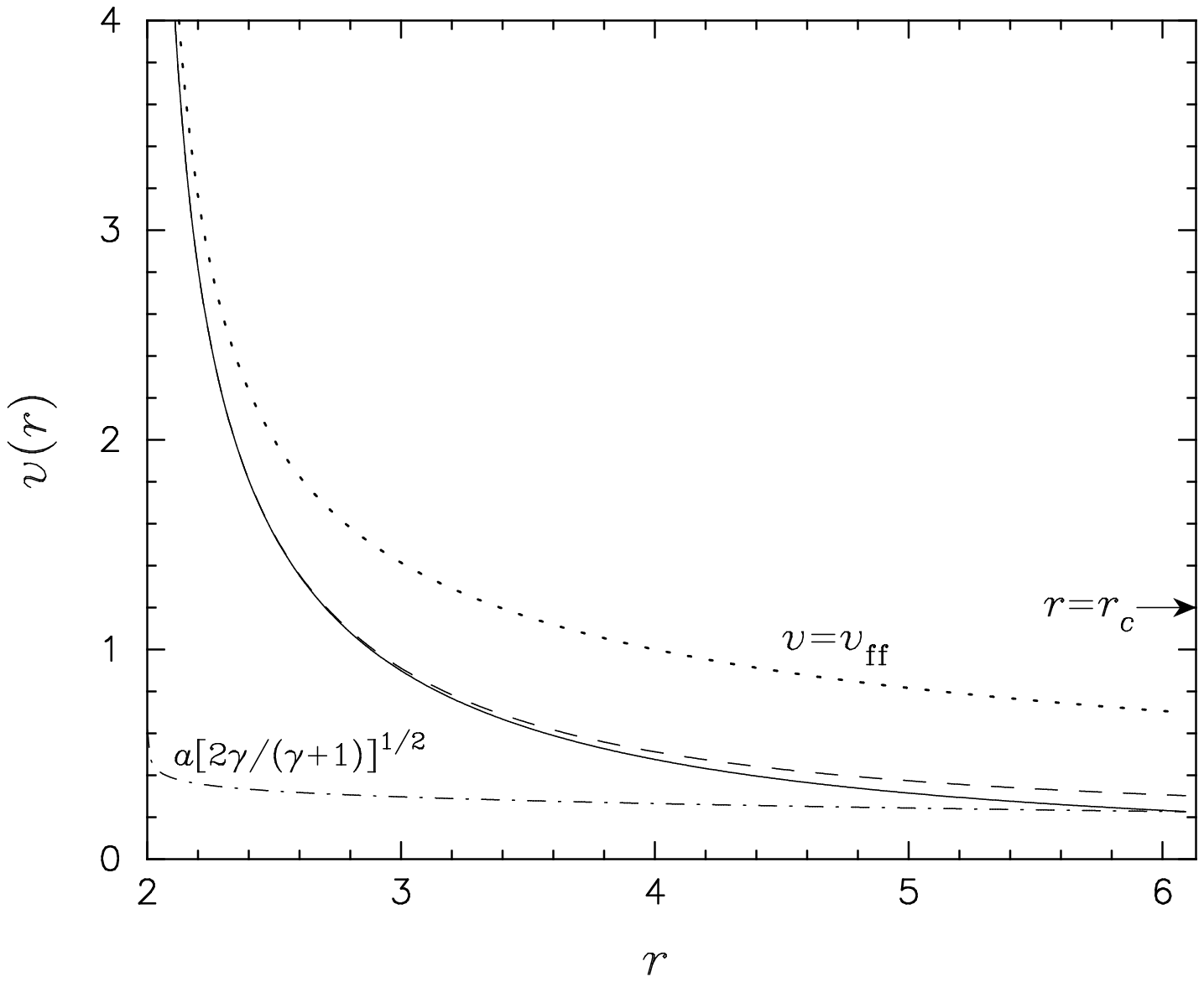}
\caption{
\baselineskip 12 true pt
Exact numerical solution for the inflow velocity $v$ ({\it solid
line}) plotted as a function of radius $r$ for Model~1, with
$\alpha = 0.1$ and $\ell_0 = 2.6$. Included for
comparison is the asymptotic solution for $v(r)$ given by eq.~(\ref{eq44})
({\it dashed line}). The two results agree closely for $2 < r < 3$,
and remain similar out to the critical point at $r_c = 6.132$. Also
shown are the free-fall velocity $\vff = [2GM/(r-\rs)]^{1/2}$ ({\it
dotted line}), and the numerical result for the sound speed $a$ multiplied
by $[2 \gamma/(\gamma+1)]^{1/2}$ ({\it dot-dashed line}). The sound speed curve
crosses the numerical solution for $v$ at the critical point.}
\end{figure}

\clearpage

\begin{figure}
\plotone{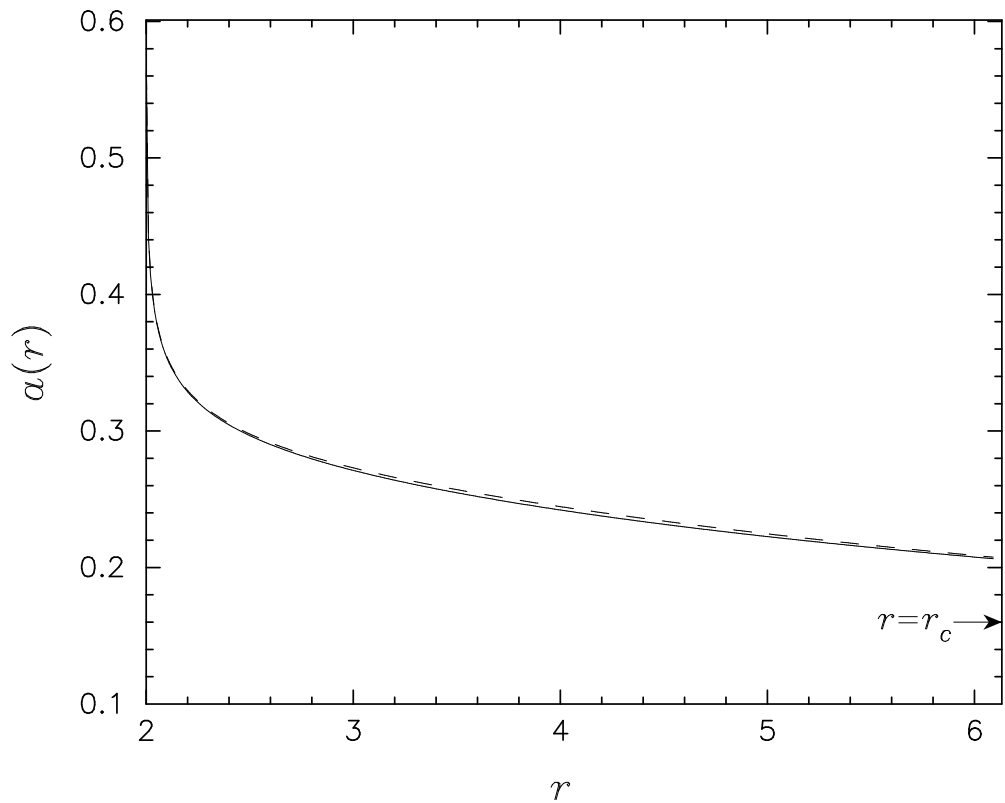}
\caption{
\baselineskip 12 true pt
Exact numerical solution for the isothermal sound speed $a$
({\it solid line}) plotted as a function of radius $r$
for Model~1. Parameters for this model are listed in Table~2.
Included for comparison is the analytical, asymptotic solution
for $a(r)$ (eq.~[\ref{eq47}]; {\it dashed line}). The two results agree
well between the starting radius $r_* = 2.001$ and the critical
point $r_c = 6.132$.}
\end{figure}

\clearpage

\begin{figure}
\plotone{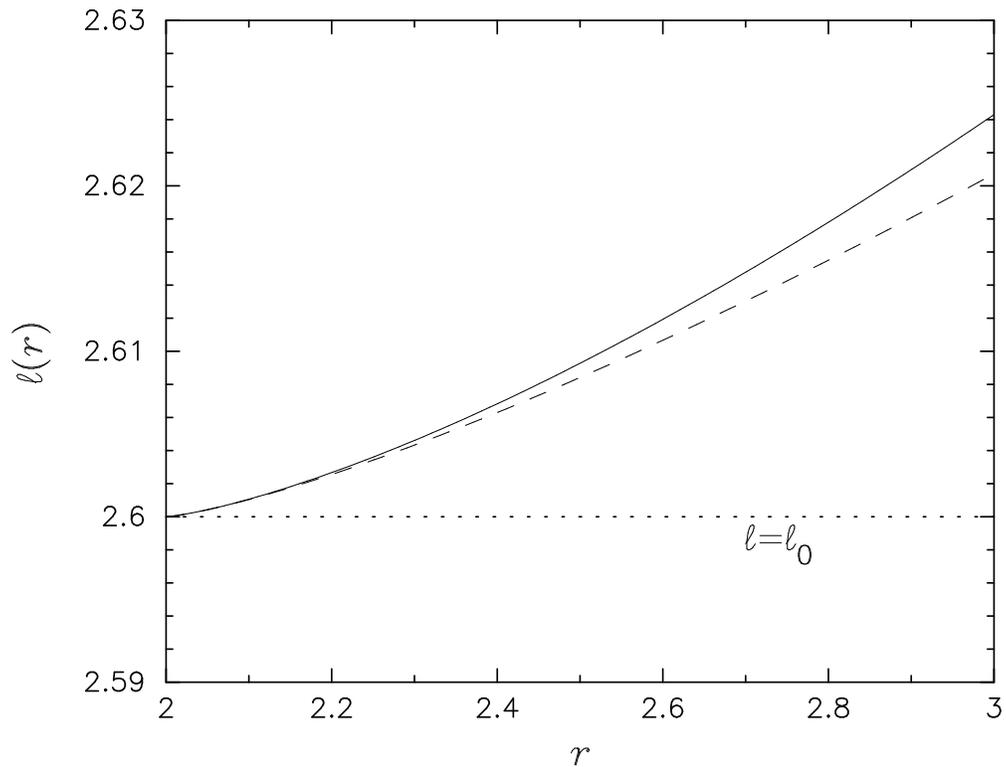}
\caption{
\baselineskip 12 true pt
Exact numerical solution for the specific angular momentum $\ell =
r^2 \, \Omega$ plotted as a function of radius $r$ ({\it solid line})
for Model~1. For comparison we include the asymptotic solution for
$\ell(r)$ given by eq.~(\ref{eq35}) ({\it dashed line}). The functions merge
smoothly as $r \to \rs$, and approach the accreted specific angular
momentum, $\ell_0 = 2.6$ ({\it dotted line}).}
\end{figure}

\clearpage

\begin{figure}
\plotone{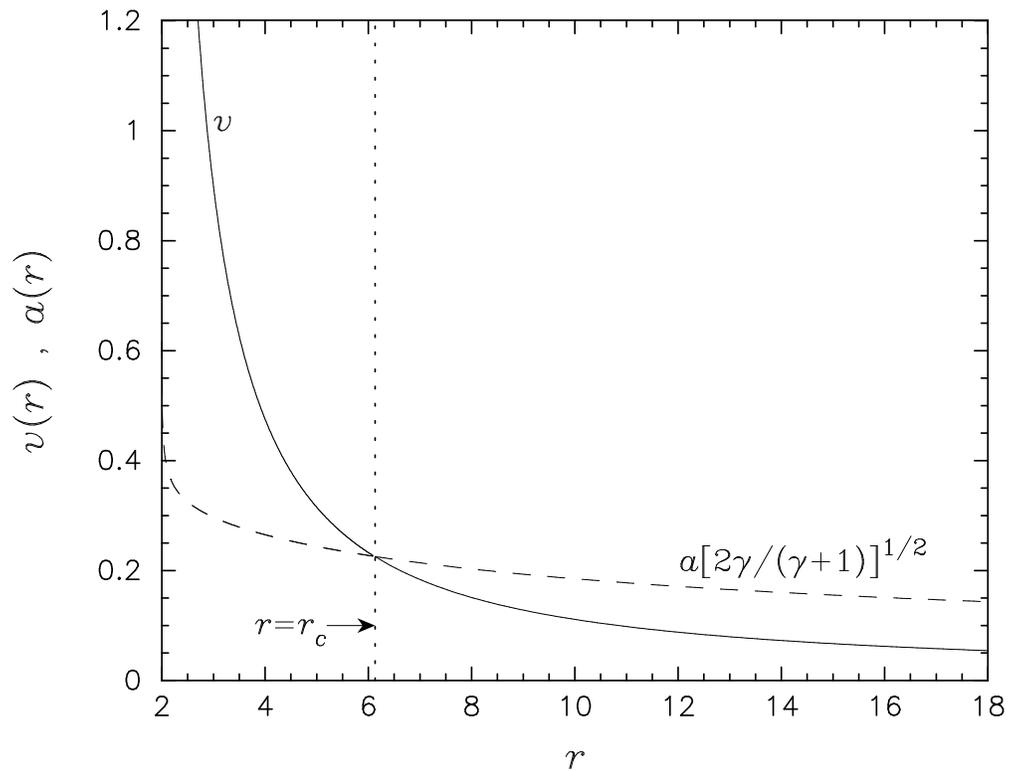}
\caption{
\baselineskip 12 true pt
Global numerical solutions for the inflow velocity $v(r)$ ({\it solid
line}) and the sound speed $a$ multiplied by $[2\gamma/(\gamma+1)]^{1/2}$
({\it dashed line}) plotted as functions of radius $r$ for Model~1.
The $v$ and $a$ curves pass smoothly through the critical point at
$r_c = 6.132$, where they cross. The flow is supersonic for $r < r_c$.
The vertical dotted line indicates the location of the critical point.}
\end{figure}

\clearpage

\begin{figure}
\plotone{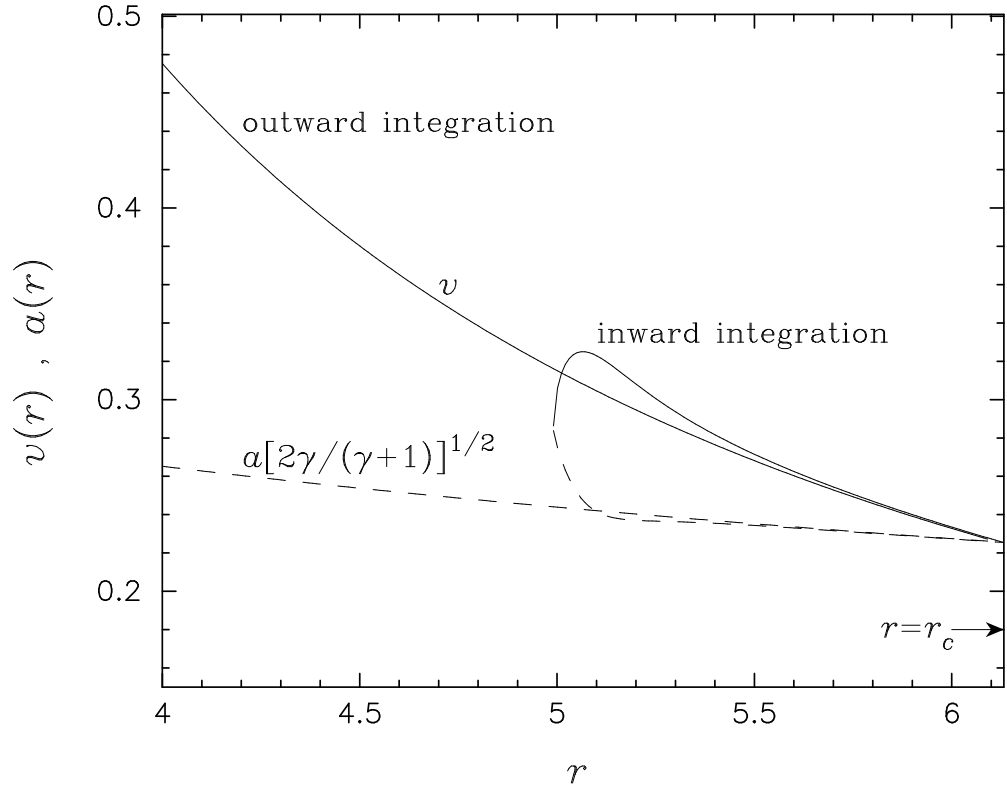}
\caption{
\baselineskip 12 true pt
Numerical solutions for the inflow velocity $v(r)$ ({\it solid line})
and the sound speed $a$ multiplied by $[2 \gamma/(\gamma+1)]^{1/2}$
({\it dashed line}) plotted as functions of radius $r$ for Model~1.
Separate results are indicated for the inwardly and outwardly directed
integrations. Note that the two sets of results agree near the critical
radius at $r_c = 6.132$. However, the inwardly-directed integration
fails at $r \sim 5$, where $v^2 = 2 \, \gamma \, a^2 / (1+\gamma)$.
See the discussion in the text.}
\end{figure}

\clearpage

\begin{figure}
\plotone{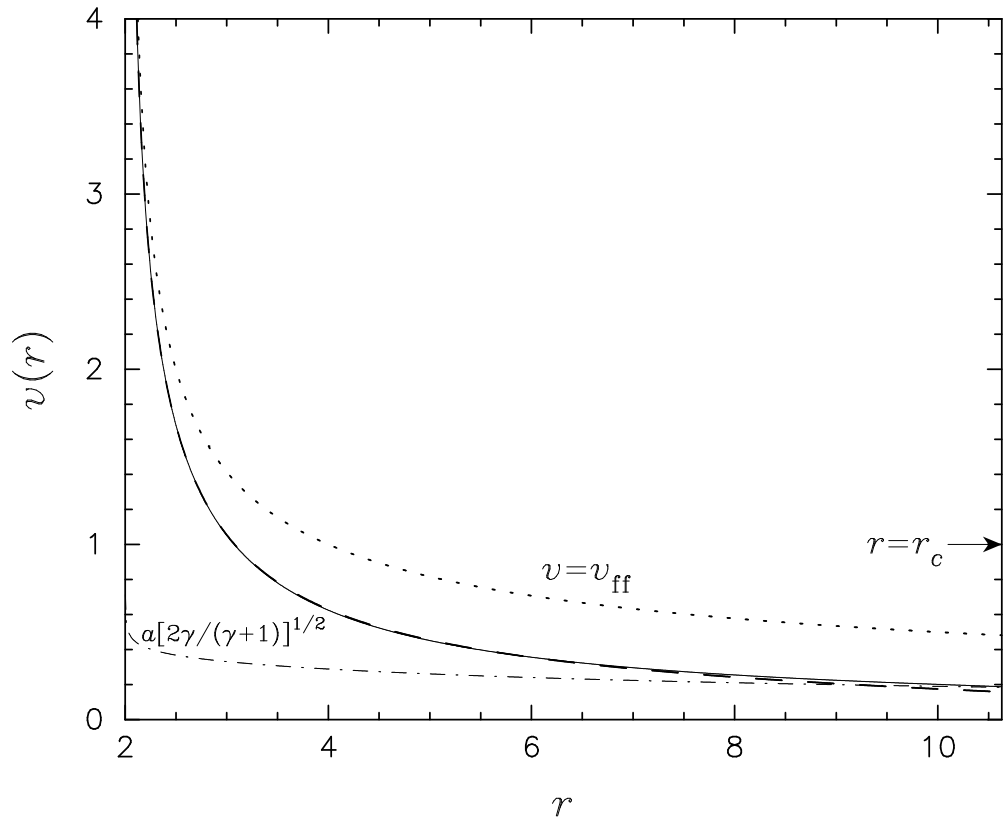}
\caption{
\baselineskip 12 true pt
Same as Fig.~1, except results correspond to Model~2, with
$\alpha = 0.3$, $\ell_0 = 1.76$, and $r_c = 10.63$. Note that
in this case, the asymptotic solution for $v(r)$ is indistinguishable from
the exact numerical solution in the entire region between the horizon and
the critical point. Additional parameters for this model are listed in Table~2.}
\end{figure}

\clearpage

\begin{figure}
\plotone{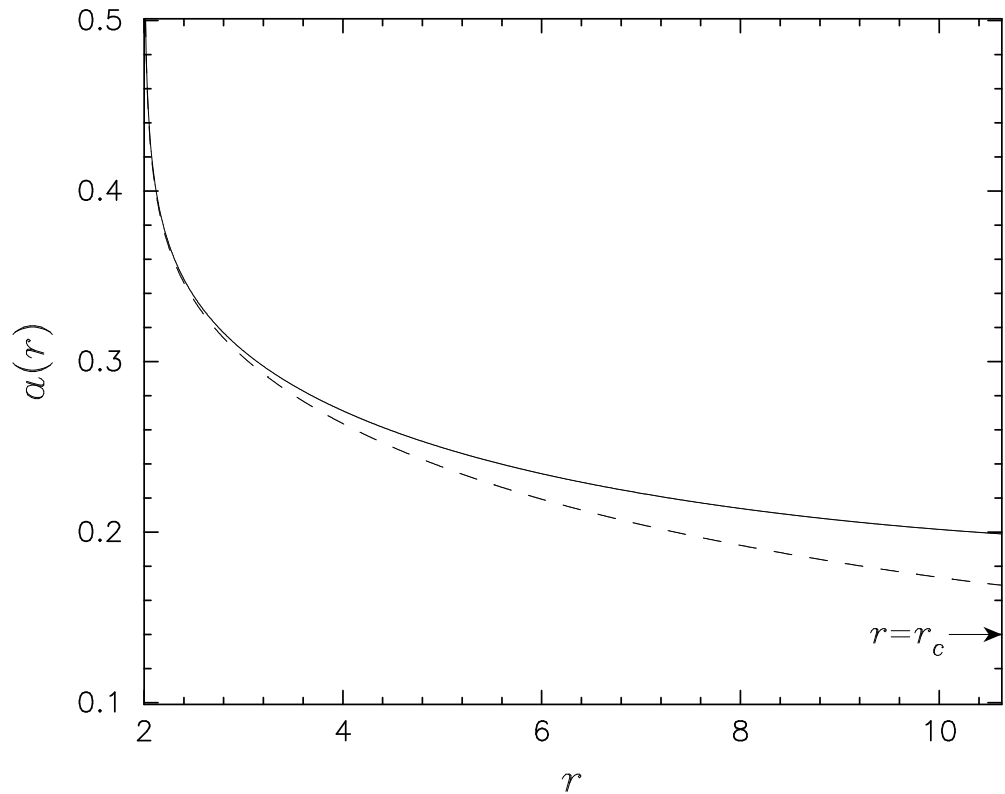}
\caption{
\baselineskip 12 true pt
Comparison of solutions for the isothermal sound speed $a(r)$ obtained
in Model~2. In this case, the exact numerical solution ({\it solid line})
agrees with the analytical solution ({\it dashed line}) close to the horizon,
but diverges for $r \gapprox 5$. The location of the critical radius at
$r_c = 10.63$ is indicated.}
\end{figure}

\clearpage

\begin{figure}
\plotone{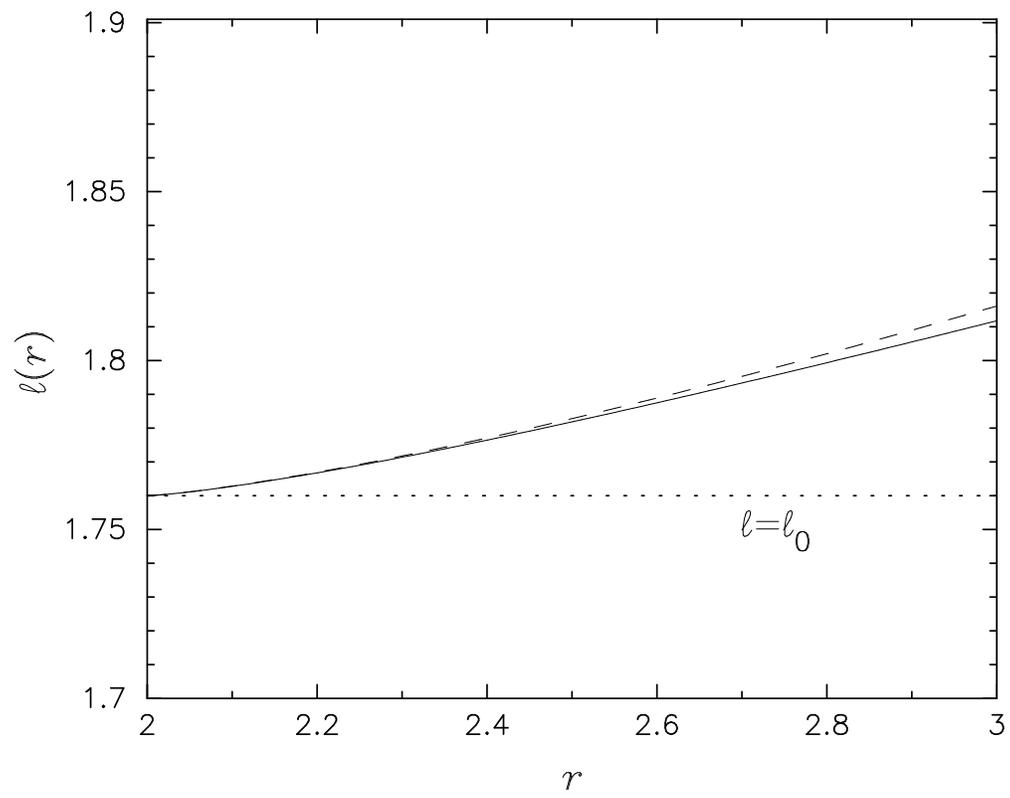}
\caption{
\baselineskip 12 true pt
Same as Fig.~3, except results correspond to Model~2 parameters.}
\end{figure}

\clearpage

\begin{figure}
\plotone{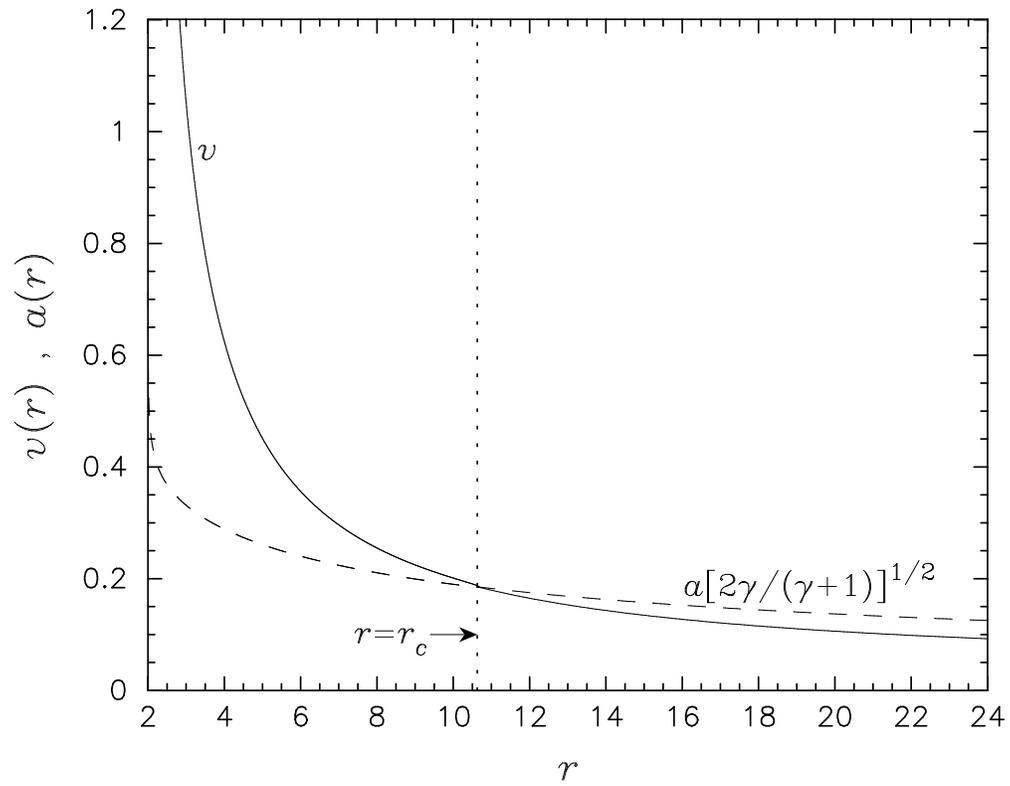}
\caption{
\baselineskip 12 true pt
Same as Fig.~4, except results correspond to Model~2 parameters. The solution
passes smoothly through a critical point located at $r_c = 10.63$.}
\end{figure}


\clearpage

\begin{deluxetable}{cccccc}
\tabletypesize{\scriptsize}
\tablecaption{Exponents of Solutions as Functions
of $\gamma$. \label{tbl-1}}
\tablewidth{0pt}
\tablehead{
\colhead{${\rm Angular \, Momentum} \atop \beta=(\gamma+5)/(2+2\gamma)$}
& \colhead{${\rm Sound \, Speed} \atop \sigma=(1-\gamma)/(2+2\gamma)$}
& \colhead{${\rm Pressure} \atop \lambda=-\gamma/(\gamma+1)$}
& \colhead{${\rm Density} \atop \eta=-1/(\gamma+1)$}
& \colhead{${\rm Disk \, Height} \atop \delta=(\gamma+3)/(2+2\gamma)$}
& \colhead{${\rm Adiabatic \, Index} \atop \gamma$}
}
\startdata
1.25 &$-$0.125 &$-$0.625 &$-$0.375 &0.875 &5/3 \\
1.27 &$-$0.115 &$-$0.615 &$-$0.385 &0.885 &8/5 \\
1.30 &$-$0.100 &$-$0.600 &$-$0.400 &0.900 &3/2 \\
1.33 &$-$0.083 &$-$0.583 &$-$0.417 &0.917 &7/5 \\
1.36 &$-$0.071 &$-$0.571 &$-$0.429 &0.929 &4/3 \\
\enddata


\tablecomments{These are the exponents of $(r-\rs)$ for the various
physical quantities close to the horizon; see the discussion in
the text.}

\end{deluxetable}


\clearpage

\begin{deluxetable}{cccccccccccccc}
\tabletypesize{\scriptsize}
\tablecaption{Model Parameters. \label{tbl-2}}
\tablewidth{0pt}
\tablehead{
\colhead{Model}
& \colhead{$\gamma$}
& \colhead{$\alpha$}
& \colhead{$\epsilon_0$}
& \colhead{$\ell_0$}
& \colhead{$K_0$}
& \colhead{$K_c$}
& \colhead{$r_c$}
& \colhead{$v_c$}
& \colhead{$a_c$}
& \colhead{$\ell_c$}
& \colhead{$r_*$}
& \colhead{$v_*$}
& \colhead{$a_*$}
}
\startdata
1
&1.5
&0.10
&0.0
&2.60
&0.007173
&0.005222
&6.132
&0.2254
&0.2058
&2.763
&2.001
&44.6812
&0.5633 \\
2
&1.5
&0.30
&0.0
&1.76
&0.014750
&0.007733
&10.63
&0.1855
&0.1694
&2.258
&2.001
&44.6843
&0.6507 \\
3
&1.5
&0.03
&0.0
&3.21
&0.001858
&0.001634
&4.900
&0.2084
&0.1902
&3.256
&2.001
&44.6802
&0.4299 \\
 \enddata

\end{deluxetable}

\end{document}